\documentclass{JHEP3}
\input epsf
\usepackage{epsfig}
\usepackage{amssymb}
\usepackage{amsmath}
\usepackage[active]{srcltx}

\newcommand{\bea}{\begin{eqnarray}}
\newcommand{\eea}{\end{eqnarray}}
\newcommand{\bean}{\begin{eqnarray*}}
\newcommand{\eean}{\end{eqnarray*}}

\def\W #1{\widetilde{#1}}

\def\ket#1{\left| #1\right\rangle}

\def\gb #1{ \left\langle #1 \right]}

\def\vev#1{\left\langle #1 \right\rangle}

\def\det{\mathop{\rm det}}

\def\a{{\alpha}}

\def\b{{\beta}}

\def\la{\lambda}
\def\eps{\epsilon}

\def\Label#1{\label{#1}}

\preprint{ITFA-2008-10}
\title{
Polynomial Structures in One-Loop Amplitudes}
\author{Ruth Britto$^{\flat}$, Bo Feng$^{\natural}$, Gang Yang$^{\sharp}$\\
~~~~\\
$^\natural$Center of Mathematical Science, Zhejiang University, Hangzhou, China\\
$^\flat$Institute for Theoretical Physics, University of Amsterdam \\
Valckenierstraat 65, 1018 XE Amsterdam, The Netherlands\\
$^\sharp$Institute of Theoretical Physics, Chinese Academy of Sciences \\
P. O. Box 2735, Beijing 100190, China }
\abstract{ A general one-loop scattering amplitude may be expanded
in terms of master integrals. The coefficients of the master
integrals can be obtained from tree-level input in a two-step
process.  First, use known formulas to write the coefficients of
$(4-2\eps)$-dimensional master integrals; these formulas depend on
an additional variable, $u$, which encodes the dimensional shift.
Second, convert the $u$-dependent coefficients of
$(4-2\eps)$-dimensional master integrals to explicit coefficients of
dimensionally shifted master integrals. This procedure requires the
initial formulas for coefficients to have polynomial dependence on
$u$. Here, we give a proof of this property in the case of massless propagators.  The proof is
constructive. Thus, as a byproduct, we produce different algebraic
 expressions for the scalar integral coefficients, in which the
 polynomial property is apparent. In these formulas, the box and
 pentagon contributions are separated explicitly.}

\begin{document}
%%%%%%%%%%%%%%%%%%%%%%%%%%%%%%%%%%%%%%%%%%%%%%%%%%%%%%%%%%%%%%%%%%

%%%%%%%%%%%%%%%%%%%%
\section{Introduction}
%%%%%%%%%%%%%%%%%%%%

Detailed calculations of multi-particle scattering events are needed in order to analyze new physics at the experiments of the Large Hadron Collider.
  Computational complexity increases
rapidly with the number of legs, even at the amplitude level.
New and improved algorithms are being developed to meet these needs.
Recent progress at next-to-leading order has been reviewed in \cite{Bern:2008ef}.

Scattering amplitudes at one-loop level can be understood in terms
of an expansion in  master integrals
\cite{IntegralRecursion,MasterIntegrals}.   The coefficients of the
master integrals may be obtained by direct reduction, or
alternatively by solving constraint equations derived from singular
structures, most notably unitarity cuts
\cite{Bern:1994zx,Bern:1994cg,Bern:1997sc,Cachazo:2004by,Bena:2004xu,Cachazo:2004dr,Britto:2004nj,Bern:2004ky,Bidder:2004tx,Britto:2004nc,Britto:2005ha,Brandhuber:2005jw,Britto:2006sj,Ossola:2006us,Anastasiou:2006jv,Mastrolia:2006ki,Britto:2006fc,Anastasiou:2006gt,Ossola:2007bb,Forde:2007mi,Ellis:2007br,BjerrumBohr:2007vu,Ossola:2007ax,Britto:2007tt,Kilgore:2007qr,Giele:2008ve,Ossola:2008xq}.
In order to obtain complete physical amplitudes from unitarity cuts,
we can work in dimensional regularization, where $D=4-2\eps$
\cite{vanNeerven:1985xr,Bern:1995db,Bern:1996je,Bern:1996ja}. By
now, explicit analytic formulas for these coefficients are available
\cite{Britto:2006fc,Forde:2007mi,Britto:2007tt,Kilgore:2007qr}.  The
input quantities are taken from the complete tree-level amplitudes
involved in unitarity cuts.  There are other promising algorithms
for finding the coefficients in 4 or $D$ dimensions
\cite{delAguila:2004nf,Ossola:2006us,Ellis:2007br,Giele:2008ve}, or
specifically the additional ``rational'' parts supplementing a pure
4-dimensional expansion
\cite{Bern:2005hs,Bern:2005ji,Bern:2005cq,Berger:2006ci,Berger:2006vq,Xiao:2006vr,Su:2006vs,Xiao:2006vt,Binoth:2006hk,Ossola:2008xq,Badger:2008cm}.

 The  formulas of \cite{Britto:2007tt}, developed in the context of
the  $D$-dimensional unitarity analysis of \cite{Anastasiou:2006jv,
Anastasiou:2006gt}, are coefficients of $(4-2\eps)$-dimensional
master integrals; these formulas depend on an additional variable,
$u$, which encodes the dimensional shift. To finish the calculation,
we  convert the $u$-dependent coefficients of
$(4-2\eps)$-dimensional master integrals to explicit coefficients of
dimensionally shifted master integrals.%
\footnote{As an alternative
to this last step, complete coefficients of $(4-2\eps)$-dimensional
master integrals could be obtained with the recursion and reduction
formulas of \cite{Anastasiou:2006jv, Anastasiou:2006gt}.}
%This procedure requires the initial formulas for coefficients to have
%polynomial dependence on $u$.   In previous work, some evidence for this assumption was provided.  Now, we give a complete proof.  As a byproduct, we offer alternate expressions for the coefficient formulas, in which the $u$-dependence has been simplified.

We are presently concerned with the adaptation of the formulas of \cite{Britto:2007tt} to an efficient numerical algorithm.  Two particular issues are addressed in this paper:
\begin{itemize}

\item Because the coefficients of the $(4-2\eps)$-dimensional integrals are polynomials in the variable $u$, a direct numerical implementation is not obvious.

\item The algebraic expression of boxes includes both box and
pentagon contributions. The pentagon contribution is signaled by the $(a u+b)$ factor in the denominator.

\end{itemize}
Our aim is to solve these two problems. More
concretely, in this paper we accomplish the following:
\begin{itemize}

\item  The proof of the polynomial property of $u$:  In previous work, some evidence for this assumption was provided.  Now, we give a complete proof.

\item  Simplifying our previous expressions: The algebraic
expressions for coefficients given in \cite{Britto:2007tt}
were the full polynomials in $u$, i.e. a sum of terms of the form $c_n u^n$.
Here, we give expressions for evaluating
$c_n$ directly from
input quantities.

\item  Separating the coefficients of boxes and pentagons: We
 give explicit, separate expressions for  coefficients of boxes and
pentagons.

\end{itemize}

For simplicity, the results here are specific to amplitudes with
massless propagators.  Generalization to the massive case is
straightforward for the coefficients of master integrals that have
nonvanishing cuts.
Based on the present paper, the generalization to the massive case has been presented in
\cite{Britto:2008vq}.  We work within the spinor formalism
\cite{SpinorFormalism}, reviewed in \cite{Dixon:2005cf}.

The paper is  organized as follows.  In Section 2, we organize our
input quantities from tree amplitudes, define some key vectors and spinors
from the input, and briefly discuss the dimensional shift.  Then we
proceed to the simplifications of the formulas for coefficients, and
the proofs that they are
polynomials in $u$.  These are given in Sections 3 and 4 for triangles and bubbles, respectively.
In Section 5, we address box coefficients, and for the first time we present separate formulas for box and pentagon coefficients.
Section 6 contains an application of these formulas, within the example of the 5-gluon amplitude.
In Section 7, we close with a discussion and comparison to a couple of other recent approaches to the problem of one-loop amplitudes.
Appendix A contains our definitions of master integrals and dimensional shift identities.
Appendix B contains alternate, more explicit expressions for the triangle coefficients which may be better suited for numerical evaluation,
since the derivatives have been taken analytically in every case that arises in a renormalizable theory.
Appendix C contains many of the details of the polynomial proof for bubble coefficients.
Appendix D contains analytic expressions used in cuts of pentagons.

%%%%%%%%%%%%%%%%%%%%%%%%%%%%
\section{Setup and Definitions}
%%%%%%%%%%%%%%%%%%%%%%%%%%%%

In this section, we set up some key conventions and definitions used in expressing the coefficients of master integrals, and in our proofs of polynomial dependence.

%%%%%%%%%%%%%%%%%
\subsection{Unitarity method}
%
%%%%%%%%%%%%%%%%%

The unitarity cut of a one-loop amplitude is its discontinuity
across a branch cut in a kinematic region selecting a particular
momentum channel.
  Specifically, we denote the momentum vector by $K$.  Then, $K^2$ should be positive, and all
other momentum invariants should be negative.  The vector $K$ will
be a sum of momenta of some of the external legs.  The discontinuity
is given by \bea
 \Delta A^{\rm 1-loop} = \int d^D\Phi~~ A^{\rm tree}_{\rm Left} ~\times~
 A^{\rm tree}_{\rm Right},
\label{cutdef} \eea where the Lorentz-invariant phase-space (LIPS)
of a double cut is defined by inserting two $\delta$-functions
representing the cut conditions:\footnote{The delta functions here
should properly be
  denoted by $\delta^{(+)}$, indicating that they are restricted to
  the positive light cone.  We shall drop the superscript for simplicity.}
\bea d^D\Phi=d^Dp~ \delta(p^2) \delta((p-K)^2) \eea

The ``unitarity method'' \cite{Bern:1994zx,Bern:1994cg}  combines
the unitarity cuts with the results of reduction to an expansion in
master integrals $I_i$ \cite{MasterIntegrals} \bea A^{\rm 1-loop} =
\sum_i c_i I_i. \label{pvred} \eea The master integrals in $d$
dimensions with massless propagators are scalar pentagons, scalar
boxes, scalar triangles, and scalar bubbles.  In the full
$d$-dimensional formalism, there are no cut-free terms.

The $n$-point scalar integral with massless propagators is
\bea  & & - i (4\pi)^{D/2} \int {d^{D} p \over (2\pi)^{D}}{1\over
p^2 (p-K)^2 \prod_{j=1}^{n-2} (p-P_j)^2}. \label{n-scalar}
 \eea

The coefficients $c_i$ in (\ref{pvred}) are, by construction,
cut-free rational functions. In the unitarity method, we do not
derive the coefficients of master integrals by performing any
reduction.  Rather, we take the coefficients as unknowns and proceed
to constrain them by performing cuts on both sides of (\ref{pvred}):
\bea \Delta A^{\rm 1-loop} = \sum_i c_i \Delta I_i. \label{method}
\eea

Any realization of the unitarity method must address the problem of
isolating the individual coefficients $c_i$. The unitarity method
succeeds because the cuts of master integrals are logarithms of
unique functions of the kinematic invariants.

In \cite{Britto:2004nc}, it was shown how to obtain scalar box
coefficients directly by cutting four propagators rather than two.
  Similarly explicit analytic formulas for the other
  coefficients have recently become available
  \cite{Britto:2006fc,Forde:2007mi,Britto:2007tt,Kilgore:2007qr}.

Here, we refer to  the formulas given in \cite{Britto:2007tt}, after
setting propagator masses to zero for simplicity.  The
generalization to the case of massive propagators has now been given
in \cite{Britto:2008vq}.

%%%%%%%%%%%%%%%%%
\subsection{Input quantities}
%%%%%%%%%%%%%%%%%

Having present the general picture, in this subsection we can start
with the following most general expression for a unitarity cut
integral:
\bea  C & = &  \int d^{4-2\eps} p~ c(\mu^2)\frac{ \prod_{i=1}^{m}
(-2 \W\ell \cdot P_i ) }{\prod_{j=1}^k (p-K_j)^2} \delta^{(+)}(p^2)
\delta^{(+)}((p-K)^2).
~~~\Label{I-inte}\eea
 We work in the
four-dimensional helicity scheme, so that all external momenta $K_i$
are $4$-dimensional and only the internal momentum $p$ is
$(4-2\eps)$-dimensional.
 We decompose the  $(4-2\eps)$-dimensional loop momentum as \cite{Bern:1995ix,Bern:1995db}
\bea
p=\W \ell+\vec{\mu},
\eea
where $\W \ell$ is $4$-dimensional and $\vec{\mu}$ is
$(-2\eps)$-dimensional. With the integrand in the form of (\ref{I-inte}),
there is a prefactor $c(\mu^2)$ which depends on the external momenta as
well as on $\mu^2$.  In this discussion we shall be paying careful
attention to all dependence on $\mu^2$.

From this starting point, the coefficients of master integrals were
listed in  \cite{Britto:2007tt}.  Now,  we would like to be able
produce the complete 4-dimensional expression, by performing the
integral over $\mu^2$ by the recursion and reduction formulas of
\cite{Anastasiou:2006jv, Anastasiou:2006gt}. To get  this complete
answer, we need to consider the  dependence of the prefactor
$c(\mu^2)$ on $\mu^2$, along with the  power of $\mu^2$ in the
coefficient formulas of \cite{Britto:2007tt}. We consider this
dependence in terms of the dimensionless parameter $u$, defined by
\bea u={4 \mu^2\over
K^2}
%,~~~~(1-2z)=\sqrt{1-u}
.~~~~\Label{about-u}
\eea
With this definition, the cut integral (\ref{I-inte}) can then be
rewritten as
\bea %&=& \frac{(4\pi)^\eps}{\Gamma(-\eps)}~\int d\mu^2 ~
%(\mu^2)^{-1-\eps} ~ c(\mu^2) \int
%\vev{\ell~d\ell}[\ell~d\ell]~(1-2z)~{ (K^2)^{n+1}\over
%\gb{\ell|K|\ell}^{n+2}} {\prod_{j=1}^{n+k} \gb{\ell|R_j|\ell}\over
%\prod_{i=1}^k \gb{\ell|Q_i|\ell}} \nonumber\\ &=&
C = \frac{(4\pi)^\eps}{\Gamma(-\eps)}~ \left( {K^2 \over 4
}\right)^{-\eps} \int_0^1 du \ u^{-1-\eps} ~ c(\mu^2) \int
\vev{\ell~d\ell}[\ell~d\ell]~\sqrt{1-u}~{ (K^2)^{n+1}\over
\gb{\ell|K|\ell}^{n+2}} {\prod_{j=1}^{n+k}
\gb{\ell|R_j(u)|\ell}\over \prod_{i=1}^k \gb{\ell|Q_i(u)|\ell}} ~,
\qquad \Label{cut-int-form} \eea

The coefficients listed in \cite{Britto:2007tt}, which are summarized
below, are the results of the four-dimensional part of the integral
(\ref{I-inte}); they are functions  of $u$.  The ``four-dimensional
cut-constructible'' part of the amplitude could be obtained by
setting $u\to 0$ in each of these coefficients, inside the expansion
of the amplitude in master integrals.  The {\em complete
$D$-dimensional}
 amplitude requires dealing with this $u$-dependence.  Here it is enough to apply the polynomial
 reduction identities given in \cite{Anastasiou:2006jv,Anastasiou:2006gt}.
 These identities assume polynomial structure of the coefficients $C(u)$, which is proven in
 the present paper, and which may also be deduced within other approaches \cite{delAguila:2004nf}. However, if we desire a result only through
 ${\cal O}(\eps^0)$,
it may be more efficient to use the dimensionally-shifted basis
discussed in  \cite{Bern:1995ix,Giele:2008ve}. We
shall return to this point in the following subsection.

From the initial expression (\ref{I-inte}), we extract all the necessary information, as follows.  First, notice the
 triplet of integers
\bea
(m, ~k, ~n \equiv m-k)~~~\Label{mkn}
\eea
 which will play an important
role.  In particular, the value of $n$ constrains the basis of
master integrals \cite{{Bern:1994zx},{Bern:1994cg}}.  If $n\leq -2$,
there are  contributions only from boxes and pentagons. If $n\geq
-1$, contributions from triangles will kick in, and finally if
$n\geq 0$, bubble contributions show up as well. This pattern is
well known from traditional reduction techniques.

Second, we use the values of $K$, $P_i$, and $K_j$ from the expression (\ref{I-inte})
to define the vectors
$Q_j, R_j $, and related important quantities, as follows: \footnote{These
  definitions apply specifically to the case with massless
  propagators.  Only a slight modification is necessary for massive
  propagators \cite{Britto:2006fc,Britto:2008vq}. }
\bea
q_j & \equiv & K_j - {K_j\cdot K\over K^2}K \\
\a_j & \equiv & {K_j^2-K_j\cdot K\over K^2} \\
p_j & \equiv &  P_j - {P_j\cdot K\over K^2}K \\
\b_j & \equiv & - {P_j\cdot K\over K^2}
\eea
\bea
Q_j(u) & \equiv &  -(\sqrt{1-u}) q_j + \a_j K,~~~
\Label{Q-q-def}\\
 &=& -(\sqrt{1-u})K_j +\left({K_j^2 \over K^2}- (1-\sqrt{1-u}) { K_j \cdot K \over K^2}\right) K ~~~\Label{Q-K-def} \\
R_j(u) & \equiv & -(\sqrt{1-u}) p_j + \b_j K ~~~ \Label{R-p-def} \\
&=& -(\sqrt{1-u})P_j
 -(1-\sqrt{1-u})  { P_j \cdot K \over K^2} K ~~~\Label{R-K-def}
\eea
One important observation is that
\bea q_j\cdot K= p_j\cdot K =0.~~~~\Label{pk-dot-K}\eea
At this point,  we wish to make a few more remarks.
\begin{itemize}

\item  The input quantities are given by $K,K_j,P_j$. From this we
can define $q_j, \a_j, p_j, \b_j$ and $Q_j(u), R_j(u)$.  We make
reference to the number of these vectors, encoded in the
triple of integers $(m,k,n)$.

\item  To simplify notation when we set $u=0$, we
will write expressions such as $Q_j(u=0)$, or just $Q_j$.

\item  The coefficients of the master integrals are polynomials in
  $u$.  In this paper, we shall find that the maximum degrees of these
  polynomials are the following.
Pentagon: 0. Box: $[(n+2)/2]$. Triangle: $[(n+1)/2]$.  Bubble:
$[n/2]$.
 Here, $[d]$ denotes the greatest integer less than or equal to $d$.

For a renormalizable theory we have $n\leq 2$; thus we have the maximum
degrees of
2 for boxes, 1 for triangles, and 1 for bubbles.  These degrees are consistent
with \cite{Ossola:2006us,Ossola:2007bb} and \cite{Giele:2008ve}.

\item  Knowing the maximum value of the  degree of the polynomial in $u$, we can then calculate
the coefficient of $u^s$ by the formula
\bea
c_s = \left. {1\over s!} {d^s C(u)\over du^s}\right|_{u\to
0},~~~\Label{c-n-exp}
\eea
so
\bea
C(u)=\sum_{s=0}^{s_{max}} \left. {1\over s!} {d^s C(u)\over
du^s}\right|_{u\to 0} u^s.
\eea
The expression (\ref{c-n-exp}) is central in this paper. Since
$c_s$ now has an expression where $u$ does not appear (as indicated
by the right-hand-side of expression (\ref{c-n-exp})),
 it can be evaluated numerically. \footnote{See \cite{Britto:2008vq} for another
   approach that is possibly more efficient.}

\end{itemize}

 {\bf Summary  of coefficients of 4-dimensional master
integrals:\\}

For the box coefficient with momenta $K,K_r,K_s$,
\bean C[Q_r,Q_s,K] & = & {(K^2)^{2+n}\over 2}\left({\prod_{j=1}^{k+n}
\gb{P_{sr,1}|R_j |P_{sr,2}}\over \gb{P_{sr,1}|K
|P_{sr,2}}^{n+2}\prod_{t=1,t\neq i,j}^k \gb{P_{sr,1}|Q_t
|P_{sr,2}}}+ \{P_{sr,1}\leftrightarrow P_{sr,2}\}
\right).
\eean
For the triangle coefficient with momenta $K,K_s$,
\bean
 C[Q_s,K] & = & { (K^2)^{1+n}\over
2}\frac{1}{(\sqrt{\Delta_s})^{n+1}}\frac{1}{(n+1)!
\vev{P_{s,1}~P_{s,2}}^{n+1}} \nonumber
\\ & & \times \frac{d^{n+1}}{d\tau^{n+1}}\left.\left({\prod_{j=1}^{k+n}
\vev{P_{s,1}-\tau P_{s,2} |R_j Q_s|P_{s,1}-\tau P_{s,2}}\over
\prod_{t=1,t\neq s}^k \vev{P_{s,1}-\tau P_{s,2}|Q_t Q_s
|P_{s,1}-\tau P_{s,2}}} + \{P_{s,1}\leftrightarrow
P_{s,2}\}\right)\right|_{\tau=0}.
\eean
For the bubble coefficient with momentum $K$,
\bean
 C[K] = (K^2)^{1+n} \sum_{q=0}^n {(-1)^q\over q!} {d^q \over
ds^q}\left.\left( {\cal B}_{n,n-q}^{(0)}(s)+\sum_{r=1}^k\sum_{a=q}^n
\left({\cal B}_{n,n-a}^{(r;a-q;1)}(s)-{\cal
B}_{n,n-a}^{(r;a-q;2)}(s)\right)\right)\right|_{s=0}
\eean
where we have made the following definitions:
\bean {\cal B}_{n,t}^{(0)}(s)\equiv {d^n\over d\tau^n}\left.\left( {1
\over n! [\eta|\W \eta K|\eta]^{n}}  {(2\eta\cdot K)^{t+1} \over
(t+1) (K^2)^{t+1}}{\prod_{j=1}^{n+k} \vev{\ell|R_j
(K+s\eta)|\ell}\over \vev{\ell~\eta}^{n+1} \prod_{p=1}^k \vev{\ell|
Q_p(K+s\eta)|\ell}}|_{\ket{\ell}\to |K-\tau \W \eta|\eta]
}\right)\right|_{\tau= 0},
\eean
\bean & & {\cal B}_{n,t}^{(r;b;1)}(s)  \equiv  {(-1)^{b+1}\over
 b! \sqrt{\Delta_r}^{b+1} \vev{P_{r,1}~P_{r,2}}^b}{d^b \over d\tau^{b}}
\left({1\over (t+1)} {\gb{P_{r,1}-\tau
P_{r,2}|\eta|P_{r,1}}^{t+1}\over \gb{P_{r,1}-\tau
P_{r,2}|K|P_{r,1}}^{t+1}}\right. \nonumber \\ & & \times
\left.\left. {\vev{P_{r,1}-\tau P_{r,2}|Q_r \eta|P_{r,1}-\tau
P_{r,2}}^{b} \prod_{j=1}^{n+k} \vev{P_{r,1}-\tau P_{r,2}|R_j
(K+s\eta)|P_{r,1}-\tau P_{r,2}}\over \vev{P_{r,1}-\tau P_{r,2}|\eta
K|P_{r,1}-\tau P_{r,2}}^{n+1} \prod_{p=1,p\neq r}^k
\vev{P_{r,1}-\tau P_{r,2}| Q_p(K+s\eta)|P_{r,1}-\tau
P_{r,2}}}\right)\right|_{\tau=0},
\eean
\bean & & {\cal B}_{n,t}^{(r;b;2)}(s)  \equiv  {(-1)^{b+1}\over
 b! \sqrt{\Delta_r}^{b+1} \vev{P_{r,1}~P_{r,2}}^{b}}{d^{b} \over d\tau^{b}}
\left({1\over (t+1)} {\gb{P_{r,2}-\tau
P_{r,1}|\eta|P_{r,2}}^{t+1}\over \gb{P_{r,2}-\tau
P_{r,1}|K|P_{r,2}}^{t+1}}\right. \nonumber \\ & & \times
\left.\left. {\vev{P_{r,2}-\tau P_{r,1}|Q_r \eta|P_{r,2}-\tau
P_{r,1}}^{b} \prod_{j=1}^{n+k} \vev{P_{r,2}-\tau P_{r,1}|R_j
(K+s\eta)|P_{r,2}-\tau P_{r,1}}\over \vev{P_{r,2}-\tau P_{r,1}|\eta
K|P_{r,2}-\tau P_{r,1}}^{n+1} \prod_{p=1,p\neq r}^k
\vev{P_{r,2}-\tau P_{r,1}| Q_p(K+s\eta)|P_{r,2}-\tau
P_{r,1}}}\right)\right|_{\tau=0}.
\eean
%. Various variables and functions will be defined later.

Note that the prefactor $c(\mu^2)$ has not been included in these
formulas for coefficients.

%%%%%%%%%%%%%%%%%%%%
\subsection{Some important constructions from input quantities}
%%%%%%%%%%%%%%%%%%%%

 Given two momenta $S,R$, we  construct two null momenta. If $R^2=0$ and $S^2=0$, $R,S$ are themselves the two null momenta.
If at least one of them is not null, for example $R^2\neq 0$, then we
construct two null momenta as follows.
\bea P_{(S,R);i}= S+ x_i R,~~~~~x_1={-2 S\cdot R
+\sqrt{\Delta(S,R)}\over 2 R^2},~~x_2={-2 S\cdot R
-\sqrt{\Delta(S,R)}\over 2 R^2},~~~\Label{P-RS}\eea
where
\bea \Delta(S,R)\equiv (2R\cdot S)^2-4 R^2
S^2.~~~\Label{Delta-def}\eea
Then, the following quantities necessarily vanish.
\bea
0=\gb{P_{(S,R);1}|S|P_{(S,R);2}}=\gb{P_{(S,R);2}|S|P_{(S,R);1}}
=\gb{P_{(S,R);1}|R|P_{(S,R);2}}=\gb{P_{(S,R);2}|R|P_{(S,R);1}}.~~~\Label{P12-0}
\eea

%Given three generic momenta $K_2, K_3, K_4$, we can construct another
%momentum $(\W q_0)_\mu^{(K_2, K_3, K_4)}$ orthogonal to all three:
%%
%\bea   (\W q_0)_\mu^{(K_2, K_3, K_4)}\equiv \eps_{\mu\nu\rho\xi}
% K_2^\nu K_3^\rho K_4^\xi,~~~~\Label{W-q-0}\eea
%%
%One important property of $\W q_0$ is that
%%
%\bea 0=(\W q_0)^{(K_2, K_3, K_4)}\cdot K_2=(\W q_0)^{(K_2, K_3,
%K_4)}\cdot K_3=(\W q_0)^{(K_2, K_3, K_4)}\cdot
%K_4,~~~\Label{W-q0-pro}\eea
%%

We shall use the following identity:
\bea & & \gb{P_1|V|P_2} \gb{P_2|W|P_1}  =  tr\left( {1-\gamma_5\over
2} \not{P_1} \not{V} \not{P_2} \not{W}\right) \nonumber
\\ & = & {1\over 2} ( (2P_1 \cdot V)(2 P_2\cdot W)+(2P_1 \cdot W)(2 P_2\cdot V)
-(2P_1\cdot P_2)(2 V \cdot W) -4i\eps_{\mu \nu\sigma \rho}P_1^\mu
V^\nu P_2^\sigma W^\rho).
\Label{eva-2}\eea

Any four-dimensional momentum $K$ can be expanded in a basis of four
other independent
momenta $K_i,K_j,K_m,K_n$ by
\bea K=a_m K_m + a_n K_n+ a_i K_i+a_j K_j,~~~\Label{4D-exp} \eea
where the coefficients are given by
\bea a_m & = & {\eps(K_i, K, K_j, K_n)\over \eps(K_i, K_m, K_j,
K_n)}, \qquad a_n={\eps(K_i, K_m, K_j, K)\over \eps(K_i, K_m, K_j, K_n)},
\nonumber \\
a_i
&=&  {\eps(K, K_m, K_j, K_n)\over \eps(K_i, K_m, K_j,
K_n)}, \qquad a_j={\eps(K_i, K_m, K, K_n)\over \eps(K_i, K_m, K_j,
K_n)},~~~\Label{a-sol}\eea
with
\bea \eps(K_1,K_2, K_3, K_4)\equiv \eps_{\mu\nu\rho\xi} K_1^\mu
K_2^\nu K_3^\rho K_4^\xi.
~~~\Label{eps-def-1}\eea
At times, we will write these coefficients in the form $a_{m}^{(K_i,K_j,K_m,K_n;K)}$ to emphasize
the related quantities.

%%%%%%%%%%%%%%%%
\subsection{The dimensionally shifted basis}
%%%%%%%%%%%%%%%%%

Since, as we shall demonstrate, our coefficients are polynomials in
$u$, we can translate this information into the dimensionally
shifted basis \cite{Bern:1995ix}. More explicitly, if we define
\bea
I_n^D[P^{\a_1}... P^{\a_m}] & = & -i  (4\pi)^{D/2} \int {d^D P
\over (2\pi)^D} {P^{\a_1}... P^{\a_m}\over (P^2-m^2) ...
((P-\sum_{i=1}^{n-1} k_i)^2-m^2)},~~~~\Label{DIN-def} \eea
then we have
\bea I_n^{(4-2\eps)}[(\mu^2)^k] & = & {\Gamma(k-\eps)\over
\Gamma(-\eps)}I_n^{(4-2\eps+2k)}[1].
\eea

There are two merits of using this basis.  First,  we can throw away
all ${\cal O}(\eps)$ contributions to make the calculation easier.
Second, we improve efficiency. To use the recursion and reduction
relations, we first calculated  all the contributions by reduction
to boxes, triangles and bubbles, and then added them up. With the
dimensionally shifted basis, this process of reduction/summation can be
done in one step, simplifying calculations. The
usefulness of this dimensionally shifted basis has been discussed in
\cite{Ossola:2007bb,Giele:2008ve}. Here, for reference, we discuss this
evaluation in Appendix \ref{dsb}.

%We recall one important fact regarding this dimensionally shifted basis \cite{Ossola:2007bb}. Consider an integral of the form
%
%\bea I^{(4-2\eps); 2f}_{m;s}=\int d^{4-2\eps} \O q { \W q^{2f}
%q_1... q_{s}\over D_0 D_1 ... D_m},~~~\O q= q+\W q. \eea
%
%For $f>0$ (a very important condition), define $d=
%2(f+1-m)+s$. Then if $d<0$, the integral is ${\cal
%O}(\eps)$ and can be neglected, but if $d\geq 0$, the integral is ${\cal
%O}(1)$ and should be kept.

%To apply this conclusion to the unitarity cut integrand (\ref{I-inte}), if
%there is a $(\mu^2)^s$ term in the prefactor with $s\geq 1$, we should
%check the quantity
%
%\bea
%d=2(s-k)+m.
%\eea
%
% If $d<0$, then the result is
%${\cal O}(\eps)$ and can be neglected.  This observation can
%help with simplification.

%%%%%%%%%%%%%%%%%%%%%%
\section{Triangle coefficients}
%%%%%%%%%%%%%%%%%%%%%%

Now that we have the necessary background information, it is simplest to start with the
coefficients of triangles. Some features of this discussion will
apply to bubbles as well.

%%%%%%%%%%%%%%
\subsection{Simplifying the formula}
%%%%%%%%%%%%%%%

We write the formula for triangle coefficients from
\cite{Britto:2007tt} in the notation of the previous section,
emphasizing $u$-dependence.
\bea & & C[Q_s(u),K] =  {(K^2)^{1+n}\over
2(\sqrt{\Delta(Q_s(u),K)(u)})^{n+1}}\frac{1}{(n+1)!
\vev{P_{(Q_s(u),K);1}(u)~P_{(Q_s(u),K);2}(u)}^{n+1}}
~~~~~\Label{beg-tri-exp}
\\ & & \quad \times \frac{d^{n+1}}{d\tau^{n+1}}\left({\prod_{j=1}^{k+n}
\vev{P_{(Q_s(u),K);1}(u)-\tau P_{(Q_s(u),K);2}(u) |R_j
(u)Q_s(u)|P_{(Q_s(u),K);1}(u)-\tau P_{(Q_s(u),K);2}(u)}\over
\prod_{t=1,t\neq s}^k \vev{P_{(Q_s(u),K);1}(u)-\tau
P_{(Q_s(u),K);2}(u)|Q_t(u) Q_s (u)|P_{(Q_s(u),K);1}(u)-\tau
P_{(Q_s(u),K);2}(u)}}\right. \nonumber \\ & & \qquad\qquad\qquad +
\{P_{(Q_s(u),K);1}(u)\leftrightarrow
P_{(Q_s(u),K);2}(u)\}\Big)\Big|_{\tau\to 0},\nonumber \eea
where the $P_{(Q_s(u),K);1,2}(u)$, as depicted in the indices, are
constructed in terms of $Q_s(u), K$, as defined in (\ref{P-RS}), and
depend on $u$. In principle we can put (\ref{beg-tri-exp}) into
(\ref{c-n-exp}) to take derivatives. However, the $u$-dependence
everywhere might be an obstacle to taking stable derivatives in
terms of $u$ in (\ref{c-n-exp}). In this subsection, we recast this
$u$-dependence in a simpler form.

Using the definition of $Q_s$ from (\ref{Q-q-def}), and the property
(\ref{pk-dot-K}), we find from  (\ref{P-RS}), (\ref{Delta-def}) that
\bean \Delta(Q_s(u),K) (u)& = & (1-u) (- 4 q_s^2 K^2), \\
x_{1,2}(u) & = & { -2 \a_s K^2 \pm
\sqrt{\Delta(Q_s(u),K)}\over 2 K^2}.\eean
When we take the square root of $\Delta(Q_s(u),K)$, there is a sign
ambiguity. It can be shown that the choice of sign does not affect
the final result.  To be explicit, we choose the minus sign here, i.e.,
\bea \sqrt{\Delta(Q_s(u),K)(u)}& = & - \sqrt{1-u} \sqrt{-
4 q_s^2 K^2}~~~\Label{Delta-tri-z} \\  x_{1,2} (u)& = & -\sqrt{1-u}
\left( {\pm \sqrt{-  q_s^2 K^2}\over K^2} \right) -\a_s
\\ &=&  -(\sqrt{1-u}) y_{1,2}-\a_s,\eea
where we have defined new scalar quantities, $y_{1,2}$, as follows:
\bea
y_{1,2} \equiv \pm {\sqrt{-  q_s^2 K^2}\over K^2}
= \pm {\sqrt{(K_s \cdot K)^2 - K_s^2 K^2}\over K^2}.
\eea
With these results, we can see that
\bea P_{(Q_s(u),K);i}(u)= -(\sqrt{1-u}) q_s+\a_s K +x_{i}(u) K = -\sqrt{1-u} (
q_s+ y_{i} K)=-(\sqrt{1-u})P_{(q_s,K);i}.~~~\Label{Delta-P12-z}\eea
The $u$-dependence has been factored out; here the null momentum
$P_{(q_s,K);i}$ does {\em not} depend on $u$, since it is constructed
from $q_s, K$--as indicated in the subscript indices.

Substituting (\ref{Delta-tri-z}) and (\ref{Delta-P12-z}) back into
(\ref{beg-tri-exp}), we find that the factor $\sqrt{1-u}$ has  cancelled
out. Thus we have
\bea & &  C[Q_s,K] =  {(K^2)^{1+n}\over 2
(-\sqrt{1-u})^{n+1}(\sqrt{\Delta(q_s,K)})^{n+1}}\frac{1}{(n+1)!
\vev{P_{(q_s,K);1}~P_{(q_s,K);2}}^{n+1}} \nonumber
\\ & & \frac{d^{n+1}}{d\tau^{n+1}}\left({\prod_{j=1}^{k+n}
\vev{P_{(q_s,K);1}-\tau P_{(q_s,K);2} |R_j(u)
Q_s(u)|P_{(q_s,K);1}-\tau P_{(q_s,K);2}}\over \prod_{t=1,t\neq s}^k
\vev{P_{(q_s,K);1}-\tau P_{(q_s,K);2}|Q_t(u) Q_s
(u)|P_{(q_s,K);1}-\tau P_{(q_s,K);2}}} +
\{P_{(q_s,K);1}\leftrightarrow P_{(q_s,K);2}\}\right)\Bigg|_{\tau\to
0}. \nonumber
\eea
To simplify further, apply the identity
$\vev{\ell| Q Q |\ell}=0$ to derive
\bea \vev{\ell|Q_t(u) Q_s(u)|\ell} & = &  \vev{\ell|(Q_t(u)-
{\a_t\over \a_s} Q_s(u)) Q_s(u)|\ell} = -\sqrt{1-u}\vev{\ell|
(q_t-{\a_t\over \a_s} q_s) Q_s(u)|\ell} ~~~\Label{QQqt}
\\ \vev{\ell|R_j(u)
Q_s(u)|\ell} & = & \vev{\ell|(R_j(u)- {\b_j\over \a_s} Q_s(u))
Q_s|\ell} = -\sqrt{1-u}\vev{\ell| (p_j-{\b_j\over \a_s} q_s)
Q_s(u)|\ell}.~~~\Label{QQpj}
\eea
If we define two more vectors $\W q_t,\W p_j$ by
\bea\W q_t=(q_t-{\a_t\over \a_s} q_s),~~~\W p_j=(p_j-{\b_j\over
\a_s} q_s),
\eea
then we  use the identities (\ref{QQqt}), (\ref{QQpj}) to conclude that
\bea
& &  C[Q_s,K]  =  {(K^2)^{1+n}\over
2(\sqrt{\Delta(q_s,K)})^{n+1}}\frac{1}{(n+1)!
\vev{P_{(q_s,K);1}~P_{(q_s,K);2}}^{n+1}} \nonumber
\\ & &  \frac{d^{n+1}}{d\tau^{n+1}}\left({\prod_{j=1}^{k+n}
\vev{P_{(q_s,K);1}-\tau P_{(q_s,K);2} |\W p_j
Q_s(u)|P_{(q_s,K);1}-\tau P_{(q_s,K);2}}\over \prod_{t=1,t\neq s}^k
\vev{P_{(q_s,K);1}-\tau P_{(q_s,K);2}|\W q_t Q_s(u)
|P_{(q_s,K);1}-\tau P_{(q_s,K);2}}}  +
\{P_{(q_s,K);1}\leftrightarrow P_{(q_s,K);2}\}\right)\Bigg|_{\tau\to
0}.~~~~~\Label{beg-tri-exp-sim} \eea\\

Compared to (\ref{beg-tri-exp}), the $u$-dependence in
(\ref{beg-tri-exp-sim}) is much simpler;  all $u$-dependence here
comes {\em only} from $Q_s(u)$. Thus, (\ref{beg-tri-exp-sim}) is
well suited for use in  (\ref{c-n-exp}).\\

{\bf How to use the formula (\ref{beg-tri-exp-sim}):}
 The
degree of this polynomial in $u$ will be seen to be $[(n+1)/2]$.
Thus we can get the corresponding coefficients by taking derivatives
in $u$ first (from $0$ to $[(n+1)/2]$, to get coefficients from each
term in the polynomial), and then setting $u=0$.

For example when $n=-1, 0$ we can set $u=0$ directly and get
\bea & &  C[Q_s,K]_{n \in \{0,-1\}}  =  {(K^2)^{n+1}\over
2(\sqrt{\Delta(q_s,K)})^{n+1}}\frac{1}{
\vev{P_{(q_s,K);1}~P_{(q_s,K);2}}^{n+1}} \nonumber
\\ & &  \frac{d^{n+1}}{d\tau^{n+1}}\left({\prod_{j=1}^{k+n}
\vev{P_{(q_s,K);1}-\tau P_{(q_s,K);2} |\W p_j Q_s|P_{(q_s,K);1}-\tau
P_{(q_s,K);2}}\over \prod_{t=1,t\neq s}^k \vev{P_{(q_s,K);1}-\tau
P_{(q_s,K);2}|\W q_t Q_s |P_{(q_s,K);1}-\tau P_{(q_s,K);2}}}  +
\{P_{(q_s,K);1}\leftrightarrow P_{(q_s,K);2}\}\right)\Bigg|_{\tau\to
0},~~~~~\Label{n=0-1}\eea
which is suitable for numerical evaluation. For $n=1,2$ the result
will take the form of a linear polynomial, $c_0+c_1 u$. To get $c_1$ we take one derivative, using
\bea
\left. {d Q_s(u)\over du} \right|_{u=0}={q_s\over 2}.
\eea

In Appendix \ref{explicit}, we have  explicit
expressions, free of derivatives, for triangle coefficients when
$n\leq 2$.

The formula (\ref{beg-tri-exp-sim}) contains $u$  in both
numerator and denominator, so it is not so obvious that
the total result is simply a polynomial in $u$.  The proof of this property is given in the next subsection.

%%%%%%%%%%%%%%%%%%%
\subsection{Proof that the triangle coefficient is a polynomial in $u$}
%%%%%%%%%%%%%%%%%%%

We start by considering two quantities that arise in our expressions,
in the course of taking derivatives:
\bea E_1 & \equiv & \vev{ P_{(q_s,K);2} |\W p_j
Q_s(u)|P_{(q_s,K);1}}+\vev{P_{(q_s,K);1} |\W p_j Q_s(u)| P_{(q_s,K);2}} \\
E_2 & \equiv & \vev{P_{(q_s,K);1}| \W p_j
Q_s(u)|P_{(q_s,K);1}}\vev{P_{(q_s,K);2}|\W p_j
Q_s(u)|P_{(q_s,K);2}}\eea
 By writing $Q_s(u)$ as a linear
combination of the $P_{(q_s,K);i}$,
\bea
Q_s(u) =  \left( -{\sqrt{1-u} \over 2} +{\a_s \over 2 y_1}\right) P_{(q_s,K);1}
-  \left( { \sqrt{1-u} \over 2} +{\a_s \over 2 y_1} \right) P_{(q_s,K);2},
\eea
and recalling that $q_s\cdot K=\W p_j\cdot K=0$,
we find that
%
%\bean E_1
%& = & \vev{ P_{(q_s,K);2} |\W p_j
%Q_s(u)|P_{(q_s,K);1}}+\vev{P_{(q_s,K);1} |\W p_j
%Q_s(u)| P_{(q_s,K);2}} \\
%& = & \vev{P_{(q_s,K);1}~P_{(q_s,K);2}}\left( (-2\W p_j\cdot q_s){
%\sqrt{1-u}{-2q_s\cdot K\over K^2}+2\a_s\over {\sqrt{\Delta(q_s,K)}\over
%K^2}} +(-2\W p_j\cdot K){ \sqrt{1-u}
% 2{q_s^2\over
%K^2} +\a_s{-2q_s\cdot K\over K^2}\over {\sqrt{\Delta(q_s,K)}\over
%K^2}}\right) \\
%& = & {\vev{P_{(q_s,K);1}~P_{(q_s,K);2}}\over
%\sqrt{\Delta(q_s,K)}} [\sqrt{1-u} ( (2 q_s\cdot K)(2\W p_j\cdot q_s)- 2
%q_s^2 (2\W p_j\cdot K))+ \a_s ((2 q_s\cdot K)(2\W p_j\cdot K)- 2 K^2
%(2\W p_j\cdot q_s))]\eean
%%
%
\bea  E_1  & = & -{  \a_s  K^2 \over
\sqrt{- q_s^2 K^2}}  (2\W p_j\cdot
q_s)\vev{P_{(q_s,K);1}~P_{(q_s,K);2}}.
~~~~\Label{E1}\eea
%
%where unwanted factor $\sqrt{1-u}=\sqrt{1-u}$ has dropped out.
All $u$-dependence has dropped out of this expression.

For $E_2$, similar manipulations show that
\bea E_2
%& = & -{1 \over 4} \vev{P_{(q_s,K);1}~P_{(q_s,K);2}}^2
% \left({K^2 \a_s^2\over  q_s^2 (1-u)}+ 1 \right)
%\gb{P_{(q_s,K);1}|\W p_j P_{(q_s,K);2}\W p_j |P_{(q_s,K);1}}
%\\
 & = &  \vev{P_{(q_s,K);1}~P_{(q_s,K);2}}^2
\left(q_s^2 \W p_j^2 -  (q_s \cdot \W p_j)^2    \right)
 \left({K^2 \a_s^2 \over  q_s^2}+  1-u \right),
~~~\Label{E2}\eea
which is a polynomial in $u$.

Now we  prove that the full expression  (\ref{beg-tri-exp-sim}) for the triangle coefficient is a polynomial in $u$.  Throughout this proof, let us abbreviate $Q_s$ by $Q$ and $P_{(q_s,K);i}$ by $P_i$.

The triangle coefficient  is given in terms of derivatives with
respect to $\tau$ on an expression where the $\tau$-dependence
appears in the factors  $\vev{P_{1}-\tau P_{2} |\W p_j
Q(u)|P_{1}-\tau P_{2}}$ (in numerator or denominator).  After taking
the derivatives, we set $\tau=0$. In this process we will produce
the following three combinations: $\vev{P_{1}|\W p_j Q(u)|P_{1}}$;
$E_1$; $E_2$. It is easy to see how the first two combinations
arise.  The third combination, $E_2$, appears, for example, in
\bean {d^2 \vev{P_{1}-\tau P_{2} |\W p_j Q(u)|P_{1}-\tau
P_{2}}\over d\tau^2}={-2\vev{P_{1}| \W p_j
Q(u)|P_{1}}\vev{P_{2}|\W p_j
Q(u)|P_{2}}\over \vev{P_{1}| \W p_j
Q(u)|P_{1}}} \eean

Consider the $\tau$-dependent factors in the denominator.
With each derivative, we effectively add one overall factor of
 $\vev{P_1- \tau
P_2|X Q(u)|P_1-\tau P_2}$ in the denominator and place one new factor, either
$(\vev{P_1- \tau P_2|X Q(u)|P_2}+ \vev{P_2|X Q(u)|P_1-\tau P_2})$ or
$(-2\vev{P_2|X Q(u)|P_2}\vev{P_1|X Q(u)|P_1})$ in the numerator. After taking $n+1$
derivatives, there are $(n+1)$ additional factors $\vev{P_1- \tau P_2|X Q(u)|P_1-\tau
P_2}$ in the denominator.  Thus, we have exactly $k+n$ factors  of $\vev{P_1|X Q(u)|P_1}$  in both numerator and denominator (after taking the $\tau \to 0$
limit). The $u$-dependence is
exactly cancelled in this part, since we have
\bea \frac{\vev{P_1|X Q(u)|P_1}}{\vev{P_1|X' Q(u)|P_1}}
=\frac{\gb{P_1|X|P_2}}{\gb{P_1|X'|P_2}}. \eea
The remaining $u$-dependence comes only through the factor $E_2$ in
the numerator. By our previous calculations (\ref{E1}) and
(\ref{E2}), we see that indeed our final expression is a polynomial
in $u$. Since every sequence of two derivatives in
(\ref{beg-tri-exp-sim}) will produce one $E_2$ factor, the degree of
the polynomial is $[(n+1)/2]$.

%%%%%%%%%%%%%%%%%%%%%%%%%%%%%%
\section{Bubble coefficients}
%%%%%%%%%%%%%%%%%%%%%%%%%%%%%%

Our proof that bubble coefficients are polynomials in $u$ is more complicated
than the one for triangle coefficients, so many of the details have been relegated to Appendix \ref{bubbleproof}.
% The basic idea of the proof is to show that terms
%corresponding to $\sqrt{1-u}$ are spurious term so their
%contributions are zero.
Here we present the simplification of the $u$-dependence in the formula for bubble coefficients.  The idea of the proof is consider the bubble coefficients are polynomials in $\sqrt{1-u}$,  and show that the odd powers vanish.

%%%%%%%%%%%%%%%%%%%
\subsection{Simplification}
%%%%%%%%%%%%%%%%%%%

 The coefficient of the
bubble integral with momentum $K$ is given by \cite{Britto:2007tt}
\bea
 C[K] = (K^2)^{1+n} \sum_{q=0}^n {(-1)^q\over q!} {d^q \over
ds^q}\left.\left( {\cal B}_{n,n-q}^{(0)}(s)+\sum_{r=1}^k\sum_{a=q}^n
\left({\cal B}_{n,n-a}^{(r;a-q;1)}(s)-{\cal
B}_{n,n-a}^{(r;a-q;2)}(s)\right)\right)\right|_{s=0},~~~~~\Label{bub-exp}
\eea
where {\small
\bea {\cal B}_{n,t}^{(0)}(s)\equiv {d^n\over d\tau^n}\left.\left( {1
\over n! [\eta|\W \eta K|\eta]^{n}}  {(2\eta\cdot K)^{t+1} \over
(t+1) (K^2)^{t+1}}{\prod_{j=1}^{n+k} \vev{\ell|R_j(u)
(K+s\eta)|\ell}\over \vev{\ell~\eta}^{n+1} \prod_{p=1}^k \vev{\ell|
Q_p(u)(K+s\eta)|\ell}}|_{\ket{\ell}\to |K-\tau \W \eta|\eta]
}\right)\right|_{\tau= 0},~~~\Label{cal-B-0}\eea
\bea & & {\cal B}_{n,t}^{(r;b;1)}(s)  \equiv  {(-1)^{b+1}\over
 b! \sqrt{\Delta(Q_r(u),K)}^{b+1} \vev{P_{(Q_r(u),K);1}(u)~P_{(Q_r(u),K);2}(u)}^b}\nonumber \\
 & & {d^b \over d\tau^{b}}
\left({1\over (t+1)} {\gb{P_{(Q_r(u),K);1}(u)-\tau
P_{(Q_r(u),K);2}(u)|\eta|P_{(Q_r(u),K);1}(u)}^{t+1}\over
\gb{P_{(Q_r(u),K);1}(u)-\tau
P_{(Q_r(u),K);2}(u)|K|P_{(Q_r(u),K);1}(u)}^{t+1}}\right. \nonumber
\\ & & \times {\vev{P_{(Q_r(u),K);1}(u)-\tau P_{(Q_r(u),K);2}(u)|Q_r
(u)\eta|P_{(Q_r(u),K);1}(u)-\tau P_{(Q_r(u),K);2}(u)}^{b}\over
\vev{P_{(Q_r(u),K);1}(u)-\tau P_{(Q_r(u),K);2}(u)|\eta
K|P_{(Q_r(u),K);1}(u)-\tau P_{(Q_r(u),K);2}(u)}^{n+1}}\nonumber \\ &
& \times \left.\left. {\prod_{j=1}^{n+k}
\vev{P_{(Q_r(u),K);1}(u)-\tau P_{(Q_r(u),K);2}(u)|R_j(u)
(K+s\eta)|P_{(Q_r(u),K);1}(u)-\tau P_{(Q_r(u),K);2}(u)}\over
\prod_{p=1,p\neq r}^k \vev{P_{(Q_r(u),K);1}(u)-\tau
P_{(Q_r(u),K);2}(u)| Q_p(u)(K+s\eta)|P_{(Q_r(u),K);1}(u)-\tau
P_{(Q_r(u),K);2}(u)}}\right)\right|_{\tau=0},~~~\Label{cal-B-r-1}\eea
\bea & & {\cal B}_{n,t}^{(r;b;2)}(s)  \equiv  {(-1)^{b+1}\over
 b! \sqrt{\Delta(Q_r(u),K)}^{b+1} \vev{P_{(Q_r(u),K);1}(u)~P_{(Q_r(u),K);2}(u)}^{b}}
 \nonumber \\ & & {d^{b} \over d\tau^{b}}
\left({1\over (t+1)} {\gb{P_{(Q_r(u),K);2}(u)-\tau
P_{(Q_r(u),K);1}(u)|\eta|P_{(Q_r(u),K);2}(u)}^{t+1}\over
\gb{P_{(Q_r(u),K);2}(u)-\tau
P_{(Q_r(u),K);1}(u)|K|P_{(Q_r(u),K);2}(u)}^{t+1}}\right. \nonumber
\\ & & \times {\vev{P_{(Q_r(u),K);2}(u)-\tau P_{(Q_r(u),K);1}(u)|Q_r(u)
\eta|P_{(Q_r(u),K);2}(u)-\tau P_{(Q_r(u),K);1}(u)}^{b}\over
\vev{P_{(Q_r(u),K);2}(u)-\tau P_{(Q_r(u),K);1}(u)|\eta
K|P_{(Q_r(u),K);2}(u)-\tau P_{(Q_r(u),K);1}(u)}^{n+1} }\nonumber
\\& & \times \left.\left.{ \prod_{j=1}^{n+k}
\vev{P_{(Q_r(u),K);2}(u)-\tau P_{(Q_r(u),K);1}(u)|R_j
(u)(K+s\eta)|P_{(Q_r(u),K);2}(u)-\tau P_{(Q_r(u),K);1}(u)}\over
\prod_{p=1,p\neq r}^k \vev{P_{(Q_r(u),K);2}(u)-\tau
P_{(Q_r(u),K);1}(u)| Q_p(u)(K+s\eta)|P_{(Q_r(u),K);2}(u)-\tau
P_{(Q_r(u),K);1}(u)}}\right)\right|_{\tau=0}.~~~\Label{cal-B-r-2}\eea
} In this expression, since $P_{(Q_r(u),K);1}(u)$ and $\Delta(Q_r(u),K)$
depend on $u$, we want to simplify the above expressions
as we did in the triangle case.
 Using (\ref{Delta-tri-z}) and
(\ref{Delta-P12-z}), we see that
we can pull out some factors of $\sqrt{1-u}$, giving \footnote{One way to see it is to choose
$\ket{P_{(Q_r(u),K);i}}=\ket{P_{(q_r,K);i}}$ and
$|P_{(Q_r(u),K);i}]=-\sqrt{1-u}|P_{(q_r,K);i}]$; the factor $-\sqrt{1-u}$
cancels out immediately.}
\bea & & {\cal B}_{n,t}^{(r;b;1)}(s,u)  \equiv  {1 \over
 b! (\sqrt{1-u})^{b+1}\sqrt{\Delta(q_r,K)}^{b+1} \vev{P_{(q_r,K);1}~P_{(q_r,K);2}}^b}\nonumber \\
& &  {d^b \over d\tau^{b}} \left({1\over (t+1)}
{\gb{P_{(q_r,K);1}-\tau P_{(q_r,K);2}|\eta|P_{(q_r,K);1}}^{t+1}\over
\gb{P_{(q_r,K);1}-\tau P_{(q_r,K);2}|K|P_{(q_r,K);1}}^{t+1}}\right.
\nonumber \\ & & \times  {\vev{P_{(q_r,K);1}-\tau
P_{(q_r,K);2}|Q_r(u) \eta|P_{(q_r,K);1}-\tau P_{(q_r,K);2}}^{b}\over
\vev{P_{(q_r,K);1}-\tau P_{(q_r,K);2}|\eta K|P_{(q_r,K);1}-\tau
P_{(q_r,K);2}}^{n+1}}\nonumber \\ & & \times
\left.\left.{\prod_{j=1}^{n+k} \vev{P_{(q_r,K);1}-\tau
P_{(q_r,K);2}|R_j (u)(K+s\eta)|P_{(q_r,K);1}-\tau
P_{(q_r,K);2}}\over
 \prod_{p=1,p\neq r}^k \vev{P_{(q_r,K);1}-\tau
P_{(q_r,K);2}| Q_p(u)(K+s\eta)|P_{(q_r,K);1}-\tau
P_{(q_r,K);2}}}\right)\right|_{\tau=0},~~~\Label{cal-B-r-1-new}\eea
Similarly, for the function ${\cal B}_{n,t}^{(r;b;2)}(s,u)$,
\bea & & {\cal B}_{n,t}^{(r;b;2)}(s,u)  \equiv  {1 \over
 b! (\sqrt{1-u})^{b+1}\sqrt{\Delta(q_r,K)}^{b+1} \vev{P_{(q_r,K);1}~P_{(q_r,K);2}}^b}\nonumber \\
& &  {d^b \over d\tau^{b}} \left({1\over (t+1)}
{\gb{P_{(q_r,K);2}-\tau P_{(q_r,K);1}|\eta|P_{(q_r,K);2}}^{t+1}\over
\gb{P_{(q_r,K);2}-\tau P_{(q_r,K);1}|K|P_{(q_r,K);1}}^{t+1}}\right.
\nonumber \\ & & \times  {\vev{P_{(q_r,K);2}-\tau
P_{(q_r,K);1}|Q_r(u) \eta|P_{(q_r,K);2}-\tau P_{(q_r,K);1}}^{b}\over
\vev{P_{(q_r,K);2}-\tau P_{(q_r,K);1}|\eta K|P_{(q_r,K);2}-\tau
P_{(q_r,K);1}}^{n+1}}\nonumber \\ & & \times
\left.\left.{\prod_{j=1}^{n+k} \vev{P_{(q_r,K);2}-\tau
P_{(q_r,K);1}|R_j (u)(K+s\eta)|P_{(q_r,K);2}-\tau
P_{(q_r,K);1}}\over
 \prod_{p=1,p\neq r}^k \vev{P_{(q_r,K);2}-\tau
P_{(q_r,K);1}| Q_p(u)(K+s\eta)|P_{(q_r,K);2}-\tau
P_{(q_r,K);1}}}\right)\right|_{\tau=0},~~~\Label{cal-B-r-2-new}\eea

So, the  bubble coefficient is now given by (\ref{bub-exp}),
(\ref{cal-B-0}), (\ref{cal-B-r-1-new}) and (\ref{cal-B-r-2-new}).
Unlike the triangle, since we now have factors of the form $\vev{\ell|Q(u)
(K+s\eta)|\ell}$ instead of $\vev{\ell|Q(u) K|\ell}$, we cannot
pull out further factors of $\sqrt{1-u}$. In these expressions, the only
$u$-dependence is coming from $R_j(u)$ and $Q_t(u)$, which is much
simpler than the original formula.

The best way to use these formulas is
similar to the triangle case.
We can see from the formulas (\ref{cal-B-0}), (\ref{cal-B-r-1-new}), (\ref{cal-B-r-2-new}) that the degree of the polynomial in $u$ is $[n/2]$. Thus we can
get the corresponding coefficients by taking derivatives with respect
to  $u$ and then setting $u=0$, as in (\ref{c-n-exp}).

%%%%%%%%%%%%%%%%%%%%%%%%%%%%%%%%%%%%%
\section{The box and pentagon coefficients}
%%%%%%%%%%%%%%%%%%%%%%%%%%%%%%%%%%%%%

Now we consider the box coefficients. This is the most complicated
part, although the formula is the simplest!
There are two reasons for the complexity.
 First, the box coefficients contain not only true box
coefficients, but also pentagon contributions, indicated by a linear factor
$(au+b)$ in the denominator. We should be able to separate the box
part from the pentagon part.
 The second reason is that the null momenta $P_{(Q_j(u),
Q_i(u));s}$   depend on $u$ in a very nontrivial way (as
$Q_j(u)+ x_a Q_i(u)$), unlike the cases of triangles and bubbles.

Given the vectors $Q_i,Q_j, K$ we can construct a vector
$ q_0^{(q_i,q_j,K)}$ orthogonal to all three:
\bea (q_0)_\mu^{(q_i,q_j,K)} &\equiv&
{1 \over K^2} \epsilon_{\mu\nu \rho \xi}
q_i^\nu q_j^\rho K^\xi \\
&=&{1 \over K^2}
 \epsilon_{\mu\nu \rho \xi} K_i^\nu K_j^\rho
K^\xi.~~~\Label{q-0} \eea
%
%%
%\bea (\W q_0)_\mu^{(q_i,q_j,K)}\equiv \epsilon_{\mu\nu \rho \xi}
%q_i^\nu q_j^\rho K^\xi=\epsilon_{\mu\nu \rho \xi} K_i^\nu K_j^\rho
%K^\xi,~~~q_0^{(q_i,q_j,K)}={\W q_0^{(q_i,q_j,K)}\over
%K^2},~~~~\Label{q-0} \eea
%%
We shall make use of our collection of orthogonality relations:
%Because
%
\bea q_i\cdot K=q_j\cdot K=q_0\cdot K=0\eea
%
%Thus if we have some $q\cdot K=0$, we know that $q$ is in
%three-dimension space and can be expanded using $q_0, q_i, q_j$.
%Furthermore, we have
%
\bea q_0\cdot q_i=q_0\cdot q_j=q_0\cdot K=0~~~~\Label{q-0-pro}\eea
Observe the fixed ordering of $q_i, q_j$ in the
definition (\ref{q-0}) of $q_0^{(q_i,q_j,K)}$.  Exchanging them leads to a
minus sign difference. The ordering is connected with our definition $P_{ji}=Q_j+x
Q_i$.

Because new features arise in different cases, we have divided our
discussion into the three cases
 $k=2$,  $k=3$, and  $k\geq 4$.

In the process of simplifying our expression, we also demonstrate that
the coefficients are truly polynomials in $u$.
Thus this is a constructive proof.

%%%%%%%%%%%%%%%%%%%%%%%%%%%
\subsection{The case $k=2$}
%%%%%%%%%%%%%%%%%%%%%%%%%%%%

In this case, there is exactly one box, whose coefficient is given by
\bean
{(K^2)^{2+n}\over 2} \left( {\prod_{s=1}^{n+2} \gb{P_{(Q_j(u),
Q_i(u));1}(u)|R_s(u)|P_{(Q_j(u), Q_i(u));2}(u)}\over \gb{P_{(Q_j(u),
Q_i(u));1}(u)|K|P_{(Q_j(u), Q_i(u));2}(u)}^{n+2}}+\{ P_{(Q_j(u),
Q_i(u));1}(u)\leftrightarrow P_{(Q_j(u), Q_i(u));2}(u)\}\right).
\eean
Later, we will reduce the other cases to this expression as well.
Now we carry out a detailed calculation to show
that the $u$-dependence is polynomial, and we find an
expression where the $u$-dependence is
easier to see.

In the numerator,  the factors  $R_s(u)$ have $u$-dependence in the  form $R_s(u)=-(\sqrt{1-u}) p_s+\b_s
K$ (here we do not assume any particular form of $\b_s$), with
$p_s\cdot K=0$.
Thus the vector $p_s$ can be expanded in a basis of the three-dimensional
vectorspace orthogonal to $K$, as follows.
\bea p_s & = & a_0^{(q_i,q_j,K; p_s)}
q_0^{(q_i,q_j,K)}+a_i^{(q_i,q_j,K; p_s)} q_i+a_j^{(q_i,q_j,K; p_s)}
q_j. ~~~\Label{ps-exp-k=2}
\eea
The coefficients in this expansion are:
\bea a_i^{(q_i,q_j,K; p_s)} & = & {  (p_s\cdot q_i) q_j^2- (p_s\cdot
q_j) (q_i\cdot q_j)\over q_i^2 q_j^2- (q_i\cdot q_j)^2}
\\a_j^{(q_i,q_j,K; p_s)} & = & {  (p_s\cdot q_j) q_i^2- (p_s\cdot
q_i) (q_i\cdot q_j)\over q_i^2 q_j^2- (q_i\cdot q_j)^2} \\
a_0^{(q_i,q_j,K; p_s)} & = & { (p_s\cdot q_0^{(q_i,q_j,K)})\over
(q_0^{(q_i,q_j,K)})^2}={\eps(p_s,q_i,q_j,K)\over
K^2(q_0^{(q_i,q_j,K)})^2}={\eps(P_s,K_i,K_j,K)\over
K^2(q_0^{(q_i,q_j,K)})^2}.~~~\Label{a-0-k=2}\eea
Using this expansion, we can write
\bea R_s(u) & = &  a_0^{(q_i,q_j,K; p_s)} (-(\sqrt{1-u})
q_0^{(q_i,q_j,K)})+a_i^{(q_i,q_j,K; p_s)} Q_i (u)+a_j^{(q_i,q_j,K;
p_s)} Q_j(u)+ \b_s^{(q_i,q_j,K;p_s)} K,~~~\Label{Rs-exp-k=2}\eea
where we have defined
\bea
\b_s^{(q_i,q_j,K;p_s)} \equiv
(\b_s-a_i^{(q_i,q_j,K; p_s)} \a_i
-a_j^{(q_i,q_j,K; p_s)} \a_j).~~~~\Label{beta-package}
\eea
Using the result (\ref{P12-0}) that
\bea \gb{P_{(Q_j(u), Q_i(u));1}(u)|Q_{i/j}(u)|P_{(Q_j(u),
Q_i(u));2}(u)}=0,\eea
we have
\bea
& & \gb{P_{(Q_j(u), Q_i(u));1}(u)|R_s(u)|P_{(Q_j(u),
Q_i(u));2}(u)} \nonumber \\
& &  =  -(\sqrt{1-u})a_0^{(q_i,q_j,K; p_s)}\gb{P_{(Q_j(u),
Q_i(u));1}(u)|q_0^{(q_i,q_j,K)}|P_{(Q_j(u), Q_i(u));2}(u)}\nonumber
\\ & & +\b_s^{(q_i,q_j,K;p_s)}\gb{P_{(Q_j(u), Q_i(u));1}(u)|K|P_{(Q_j(u),
Q_i(u));2}(u)}.
\eea
Thus, we can write
\bea & & {\prod_{s=1}^{n+2} \gb{P_{(Q_j(u),
Q_i(u));1}(u)|R_s(u)|P_{(Q_j(u), Q_i(u));2}(u)}\over \gb{P_{(Q_j(u),
Q_i(u));1}(u)|K|P_{(Q_j(u), Q_i(u));2}(u)}^{n+2}}
~~~~\Label{box-u-perm}
\\
& = & \sum_{h=0}^{n+2} C_{h}^{(q_i,q_j,K)}
{(-\sqrt{1-u})^{h}\gb{P_{(Q_j(u),
Q_i(u));1}(u)|q_0^{(q_i,q_j,K)}|P_{(Q_j(u),
Q_i(u));2}(u)}^{h}\over \gb{P_{(Q_j(u),
Q_i(u));1}(u)|K|P_{(Q_j(u), Q_i(u));2}(u)}^{h}}.
\nonumber
\eea
Here we have defined
\bea  C_{h}^{(q_i,q_j,K)} =
\sum_{\stackrel{S \subseteq \{1,2,\ldots,n+2\}}{|S|=h}}
\prod_{s \in S} a_0^{(q_i,q_j,K;p_s)}
\prod_{s \in S^c}\b_s^{(q_i,q_j,K;p_s)},
\eea
where $S^c$ denotes the complement of $S$: $S^c \equiv \{1,2,\ldots,n+2\}\backslash S$.

Now we can show that the box coefficients are indeed polynomials in $u$,
in this  case where $k=2$.  By the above expansion, the coefficients are
given by sum of the following typical terms. (To simplify the formulas in this proof, we shall now write
 $P_{ji,a}(u)$  in place of
$P_{(Q_j(u), Q_i(u));a}(u)$.)
\bea & & {(-\sqrt{1-u})^h\gb{P_{ji,1}(u)|q_0|P_{ji,2}(u)}^h\over
\gb{P_{ji,1}(u)|K|P_{ji,2}(u)}^h}+{(-\sqrt{1-u})^h\gb{P_{ji,2}(u)|q_0|P_{ji,1}(u)}^h\over
\gb{P_{ji,2}(u)|K|P_{ji,1}(u)}^h}~~~~ \Label{boxpiece} \\
& = &(-\sqrt{1-u})^h
{\gb{P_{ji,1}(u)|q_0|P_{ji,2}(u)}^h\gb{P_{ji,2}(u)|K|P_{ji,1}(u)}^h
+\gb{P_{ji,2}(u)|q_0|P_{ji,1}(u)}^h\gb{P_{ji,1}(u)|K|P_{ji,2}(u)}^h\over
\gb{P_{ji,1}(u)|K|P_{ji,2}(u)}^h\gb{P_{ji,2}(u)|K|P_{ji,1}(u)}^h} \nonumber \\
& = & { (-\sqrt{1-u})^h \left[ \left(2i (1-u){\sqrt{\Delta(Q_i(u),Q_j(u))}}
  q_0^2 K^2\right)^h+   \left(-2i
(1-u){\sqrt{\Delta(Q_i(u),Q_j(u))}} q_0^2 K^2 \right)^h \right] \over \left({K^2} (1-u)^2[ (2q_i\cdot
q_j)^2-4 q_i^2 q_j^2]\right)^h }. ~~~\Label{k=2-box-S} \eea
We have used the following results (making repeated use of the identity (\ref{eva-2})):
\bean \gb{P_{ji,1}(u)|K|P_{ji,2}(u)}\gb{P_{ji,2}(u)|K|P_{ji,1}(u)} &
= &  {K^2\over Q_i(u)^2} (1-u)^2[ (2q_i\cdot q_j)^2-4 q_i^2
q_j^2]\eean
\bean \gb{P_{ji,1}(u)|q_0|P_{ji,2}(u)}\gb{P_{ji,2}(u)|K|P_{ji,1}(u)}
% & = &  -2i \eps(P_{ji,1}(u), q_0,P_{ji,2}(u) , K)
&=& 2i
(1-u){\sqrt{\Delta(Q_i(u),Q_j(u))}\over Q_i(u)^2} { q_0^2 K^2}\\
\gb{P_{ji,1}(u)|K|P_{ji,2}(u)}\gb{P_{ji,2}(u)|q_0|P_{ji,1}(u)}
%& = &-2i \eps(P_{ji,1}(u), K,P_{ji,2}(u) , q_0)
&=& -2i
(1-u){\sqrt{\Delta(Q_i(u),Q_j(u))}\over Q_i(u)^2} { q_0^2 K^2}
\eean
and
%%
%\bean \eps(P_1, q_0, P_2, K)
%& = & \eps( Q_j+ x_1 Q_i, q_0, Q_j+x _2
%Q_i, K) = (x_1-x_2) \eps( Q_i, q_0, Q_j, K) \\
%& = &
%-(1-u){\sqrt{\Delta(Q_i(u),Q_j(u)}\over Q_i(u)^2}{ \W q_0^2\over
%K^2}\eean
%%
%where
%
\bea \Delta(Q_i(u),Q_j(u)) & = & (1-u)\left\{(1-u)[ (2q_i\cdot
q_j)^2-4 q_i^2 q_j^2]+4K^2[ \a_i\a_j(2q_i\cdot q_j)-\a_i^2
q_j^2-\a_j^2 q_i^2]\right\}. ~~~~\Label{delta-box-u}\eea
Equation (\ref{k=2-box-S}) is our most important result in this
subsection. There we can see that when $h$ is odd, we get
zero.  When $h$ is even, the factor $(1-u)^{2h}$  in the denominator
is cancelled by corresponding factors in the numerator. (Notice the
overall factor of $(1-u)$ within $\Delta(Q_i(u),Q_j(u))$.)
 Thus we see that indeed the expression (\ref{boxpiece}) is a
polynomial in $u$, specifically,
\bea
\left\{
\begin{array}{ll}
 { 2 (2i)^{h} (q_0^2)^{h}\left\{(1-u)[ (2q_i\cdot q_j)^2-4
q_i^2 q_j^2]+ 4K^2[ \a_i\a_j(2q_i\cdot q_j)-\a_i^2 q_j^2-\a_j^2
q_i^2]\right\}^{h/2} \over  [ (2q_i\cdot q_j)^2-4 q_i^2
q_j^2]^{h}}
& {\rm for}~ h~ {\rm even}, \\
0 & {\rm for}~ h~ {\rm odd}.
\end{array}
\right.~~~\Label{middle-1}
\eea
It is clear that the maximum degree of the polynomial is $[{n+2
    \over 2}]$.

%%%%%%%%%%%%%%%%%%%%
\subsubsection{A simpler expression for $k=2$}
%%%%%%%%%%%%%%%%%%%%

The aim of this sub-subsection is to find another expression with the
same value as (\ref{middle-1}), but with more transparent
$u$-dependence. We have just seen that  all $u$-dependence has cancelled out, except for the
second factor of $(1-u)$ in the first term of $\Delta_{ij}(u)$, as it
appears in (\ref{delta-box-u}).
Since (\ref{middle-1}) is now an expression in terms of scalar
quantities, we can consider the effect of setting $u=0$ at the
beginning of the calculation.  Recovering the expression (\ref{middle-1})
then requires the following single modification:
{\large
\bea q_0^{(q_i,q_j,K)}\to \a^{(q_i,q_j)}(u)
q_0^{(q_i,q_j,K)},~~~\a^{(q_i,q_j)}(u)= {\sqrt{1-u+ {4K^2[
\a_i\a_j(2q_i\cdot q_j)-\a_i^2 q_j^2-\a_j^2 q_i^2]\over (2q_i\cdot
q_j)^2-4 q_i^2 q_j^2}}\over \sqrt{1+ {4K^2[ \a_i\a_j(2q_i\cdot
q_j)-\a_i^2 q_j^2-\a_j^2 q_i^2]\over (2q_i\cdot q_j)^2-4 q_i^2
q_j^2}}}.
\eea
}
Then,
\bea & & {(-\sqrt{1-u})^h\gb{P_{ji,1}(u)|q_0|P_{ji,2}(u)}^h\over
\gb{P_{ji,1}(u)|K|P_{ji,2}(u)}^h}+{(-\sqrt{1-u})^h\gb{P_{ji,2}(u)|q_0|P_{ji,1}(u)}^h\over
\gb{P_{ji,2}(u)|K|P_{ji,1}(u)}^h}\nonumber \\ & = &
{\gb{P_{ji,1}(u=0)|\a^{(q_i,q_j)}(u)(-q_0)|P_{ji,2}(u=0)}^h\over
\gb{P_{ji,1}(u=0)|K|P_{ji,2}(u=0)}^h}+{\gb{P_{ji,2}(u=0)|\a^{(q_i,q_j)}(u)(-q_0)|P_{ji,1}(u=0)}^h\over
\gb{P_{ji,2}(u=0)|K|P_{ji,1}(u=0)}^h}.
\eea
Now,
all of the $u$-dependence is concentrated within the $\a^{(q_i,q_j)}(u)$. Going back
to (\ref{box-u-perm}), we can perform a similar operation:
\bea & & {\prod_{s=1}^{n+2}\gb{P_{ji,1}(u)|R_s(u)|P_{ji,2}(u)}\over
\gb{P_{ji,1}(u)|K|P_{ji,2}(u)}^{n+2}}+\{ P_{ji,1}(u)\leftrightarrow
P_{ji,2}(u)\}\nonumber \\ &= &
{\prod_{s=1}^{n+2}\gb{P_{ji,1}(u=0)|{\W R}_s(u)|P_{ji,2}(u=0)}\over
\gb{P_{ji,1}(u=0)|K|P_{ji,2}(u=0)}^{n+2}}+\{
P_{ji,1}(u=0)\leftrightarrow
P_{ji,2}(u=0)\}.
~~~\Label{k=2-box-equiv}\eea
where we have defined
\bea {\W R}_s(u) \equiv a_0^{(q_i,q_j,K; p_s)}
\a^{(q_i,q_j)}(u)(-q_0)+\b_s^{(q_i,q_j,K;p_s)} K.
~~~~\Label{wrs-k2}
\eea
With the definition (\ref{beta-package}), we can rewrite $ {\W R}_s(u)$ as
\bean \W R_s
%& = & a_0^{(q_i,q_j,K; p_s)}
%\a^{(q_i,q_j)}(u)(-q_0)+(\b_s-a_i^{(q_i,q_j,K; p_s)} \a_i
%-a_j^{(q_i,q_j,K; p_s)} \a_j)K \\
& = & a_0^{(q_i,q_j,K; p_s)} (\a^{(q_i,q_j)}(u)-1)(-q_0)  -
a_i^{(q_i,q_j,K; p_s)} Q_i(u=0)-a_j^{(q_i,q_j,K; p_s)} Q_j(u=0) +
R_s(u=0). \eean
Now we make use of the properties
\bean
\gb{P_{ji,1}(u=0)|Q_i(u=0)|P_{ji,2}(u=0)}=\gb{P_{ji,1}(u=0)|Q_j(u=0)|P_{ji,2}(u=0)}=0,\eean
and the fact that $\W R_s(u)$ is defined such that the equation
(\ref{k=2-box-equiv}) is satisfied, we can drop the terms with $Q_{i/j}(u=0)$, so that
\bea \W R_s(u)  \equiv {p_s\cdot q_0^{(q_i,q_j,K)}\over
(q_0^{(q_i,q_j,K)})^2}
(\a^{(q_i,q_j)}(u)-1)(-q_0^{(q_i,q_j,K)})+R_s(u=0).
~~~\Label{W-R-S}
\eea
Equations (\ref{k=2-box-equiv}) and (\ref{W-R-S}) are our final
simplest result.  All $u$-dependence has been packaged inside
$(\a^{(q_i,q_j)}(u)-1)$, which is zero when $u=0$. Also, it has now
become clear that  the  degree of the polynomial in $u$ is
$[(n+2)/2]$.

{\bf Summary:} The box coefficient for $k=2$ is given by
\bea C[K_i,K_j]_{k=2} & = & {(K^2)^{2+n}\over 2} \left(
{\prod_{s=1}^{n+2} \gb{P_{(Q_j, Q_i);1}|\W R_s(u)|P_{(Q_j,
Q_i);2}}\over \gb{P_{(Q_j, Q_i);1}|K|P_{(Q_j, Q_i);2}}^{n+2}}+\{
P_{(Q_j, Q_i);1}\leftrightarrow P_{(Q_j,
Q_i);2}\}\right),~~\Label{k=2-final} \eea
with
\bea \W R_s(u) ={p_s\cdot q_0^{(q_i,q_j,K)}\over
(q_0^{(q_i,q_j,K)})^2}
(\a^{(q_i,q_j)}(u)-1)(-q_0^{(q_i,q_j,K)})+R_s(u=0).~~~\Label{W-R-S-2}
\eea
Let us emphasize again that $P_{(Q_j, Q_i);a}$ is constructed from
$Q_j(u=0)+x_a Q_i(u=0)$, so it is {\em independent} of $u$.

%%%%%%%%%%%%%%%%%%%%%%
\subsection{The case $k=3$}
%%%%%%%%%%%%%%%%%%%%%%

In this case, we will see the pentagon show up, and we shall learn how
to separate boxes from pentagons.
Again, we shall abbreviate the notation of the vector $P_{(Q_j, Q_i);a}$
by $P_{ji,a}$.
We evaluate an expression of the form:
\bea {\prod_{s=1}^{n+3} \gb{P_{ji,1}(u)|R_s(u)|P_{ji,2}(u)}\over
\gb{P_{ji,1}(u)|K|P_{ji,2}(u)}^{n+2}
\gb{P_{ji,1}(u)|Q_t(u)|P_{ji,2}(u)}}+\{ P_{ji,1}(u)\leftrightarrow
P_{ji,2}(u)\}.
\eea
Again, we would like to expand $R_s(u)$, or equivalently $p_s$, in a
suitable basis of the vectorspace orthogonal to $K$.  In this case, we
do not need to construct the vector $q_0^{(q_i,q_j,K)}$, because we now have three
vectors  $q_i, q_j, q_t$ available already.
\bea p_s & = & a_t^{(q_i,q_j,q_t; p_s)} q_t+a_i^{(q_i,q_j,q_t; p_s)}
q_i+a_j^{(q_i,q_j,q_t; p_s)} q_j.~~~\Label{ps-exp-k=3}\eea
Thus,
\bea \gb{P_{ji,1}(u)|R_s(u)|P_{ji,2}(u)} & = & a_t^{(q_i,q_j,q_t; p_s)}
\gb{P_{ji,1}(u)|Q_t(u)|P_{ji,2}(u)} +\b_s^{(q_i,q_j,q_t;
p_s)}\gb{P_{ji,1}(u)|K|P_{ji,2}(u)},~~~\Label{Rs-exp-k=3}\eea
where we have defined
\bea\b_s^{(q_i,q_j,q_t; p_s)}\equiv (\b_s-\sum_{h=i,j,k}
a_h^{(q_i,q_j,q_t; p_s)}\a_h).~~~~\Label{big-beta-k-3}
\eea
Now we can expand the rational function within the coefficient formula:
\bea & & {\prod_{s=1}^{n+3}\gb{P_{ji,1}(u)|R_s(u)|P_{ji,2}(u)}\over
\gb{P_{ji,1}(u)|K|P_{ji,2}(u)}^{n+2}\gb{P_{ji,1}(u)|Q_t(u)|P_{ji,2}(u)}}
\nonumber  \\
&& =  \sum_{h=0}^{n+3} C_h^{(q_i,q_j,q_t)}
{\gb{P_{ji,1}(u)|Q_t(u)|P_{ji,2}(u)}^{h}\over
\gb{P_{ji,1}(u)|K|P_{ji,2}(u)}^{h-1}\gb{P_{ji,1}(u)|Q_t(u)|P_{ji,2}(u)}},
~~~~\Label{with-pent}
\eea
where we have defined
\bean  C_h^{(q_i,q_j,q_t)} = \sum_{\stackrel{S \subseteq \{1,2,\ldots,n+3\}}{|S|=h}}
\prod_{s \in S} a_t^{(q_i,q_j,q_t; p_s)}
\prod_{s \in S^c}\b_s^{(q_i,q_j,q_t;p_s)}.
\eean
Now, break the sum (\ref{with-pent}) into two parts, by separating the
term with $h=0$ from the rest.  In every term with $h > 0$, the
factor $\gb{P_{ji,1}(u)|Q_t(u)|P_{ji,2}(u)}$ from the denominator is cancelled by the
numerator.  What remains is a term in the form we considered in the
previous subsection, the case $k=2$.  That part contributes only to boxes.
We shall return to that part in a moment, to find the exact box
contribution.  Notice that we can observe at this point, from the
comparison to the $k=2$ case, that the degree of the polynomial is
again $[(n+2)/2]$.

The term in (\ref{with-pent}) with $h=0$ is
\bean  \prod_{s=1}^{n+3} \b_s^{(q_i,q_j,q_t;p_s)}{
\gb{P_{ji,1}(u)|K|P_{ji,2}(u)}\over \gb{P_{ji,1}(u)|Q_t(u)|P_{ji,2}(u)}
}.
\eean
The cut of a pentagon integral has been analyzed in
\cite{Anastasiou:2006gt}.  We clarify its $u$-dependent behavior in Appendix \ref{ap:pentagon}.  It is directly related to the sum of three
cut-boxes.  The part of the cut-pentagon that is related to the
cut-box $C[Q_i,Q_j]$ is
\bea {1\over 2K^2} \left( { \gb{P_{ji,1}(u)|K|P_{ji,2}(u)}\over
\gb{P_{ji,1}(u)|Q_t(u)|P_{ji,2}(u)} }+\{ P_{ji,1}(u)\leftrightarrow
P_{ji,2}(u)\}\right).
\eea
Thus, we see that this $h=0$ term is exactly a pentagon contribution.
The coefficient of the pentagon integral must
be
\bea C[Q_i, Q_j, Q_t] = (K^2)^{3+n} \prod_{s=1}^{n+3}
\b_s^{(q_i,q_j,q_t;p_s)},~~~\Label{Pen-coeff}\eea
which is entirely independent of $u$.

Now let us return to the box coefficients, using our result
(\ref{k=2-final}), along with the definition (\ref{W-R-S}) applied
here to the vector $Q_t(u)$ instead of $R_s(u)$.
\bean & & \sum_{h=1}^{n+3} C_h^{(q_i,q_j,q_t)}
{\gb{P_{ji,1}(u)|Q_t(u)|P_{ji,2}(u)}^{h-1}\over
\gb{P_{ji,1}(u)|K|P_{ji,2}(u)}^{h-1}} \\
 & \longrightarrow & \sum_{h=1}^{n+3} C_h^{(q_i,q_j,q_t)}
{\gb{P_{ji,1}(u=0)|\W Q_t(u)|P_{ji,2}(u=0)}^{h-1}\over
\gb{P_{ji,1}(u=0)|K|P_{ji,2}(u=0)}^{h-1}} \\
& = & \sum_{h=0}^{n+3} C_h^{(q_i,q_j,q_t)}
{\gb{P_{ji,1}(u=0)|\W Q_t(u)|P_{ji,2}(u=0)}^{h}\over
\gb{P_{ji,1}(u=0)|K|P_{ji,2}(u=0)}^{h-1}\gb{P_{ji,1}(u=0)|\W
Q_t(u)|P_{ji,2}(u=0)}}\\ & & - \prod_{s=1}^{n+3}
\b_s^{(q_i,q_j,q_t;p_s)}{ \gb{P_{ji,1}(u=0)|K|P_{ji,2}(u=0)}\over
\gb{P_{ji,1}(u=0)|\W Q_t(u)|P_{ji,2}(u=0)} }\\
 & = & {\prod_{s=1}^{n+3}\gb{P_{ji,1}(u=0)|\W {\W R}_s(u)|P_{ji,2}(u=0)}\over
\gb{P_{ji,1}(u=0)|K|P_{ji,2}(u=0)}^{n+2}\gb{P_{ji,1}(u=0)|\W
Q_t(u)|P_{ji,2}(u=0)}}\\ & & - \prod_{s=1}^{n+3}
\b_s^{(q_i,q_j,q_t;p_s)}{ \gb{P_{ji,1}(u=0)|K|P_{ji,2}(u=0)}\over
\gb{P_{ji,1}(u=0)|\W Q_t(u)|P_{ji,2}(u=0)} }\eean
In the last line of this equation, we have defined
\bea \W {\W R}_s (u) &  \equiv & a_t^{(q_i,q_j,q_t; p_s)} \W Q_t
+\b_s^{(q_i,q_j,q_t; p_s)}K  ~~~~\Label{wwr-k3} \\  & = & a_t^{(q_i,q_j,q_t; p_s)}
a_0^{(q_i,q_j,K; q_t)} (\a^{(q_i,q_j)}(u)-1)(-q_0) + R_s(u=0)-
\sum_{\gamma=i,j} a_\gamma^{(q_i,q_j,q_t; p_s)} Q_\gamma(u=0).
\nonumber
\eea
The cumbersome double-tilde notation is a temporary inconvenience, to
avoid confusion with ${\W R}_s (u)$, which was defined in (\ref{wrs-k2}) for the case $k=2$.
In fact, we shall discover later that the two quantities are identical.

Using the property $\gb{P_{ji,1}(u=0)|Q_{i/j}(u=0)|P_{ji,2}(u=0)}=0$, we can
redefine this vector as follows:
\bea \W {\W R}_s(u) \equiv a_t^{(q_i,q_j,q_t; p_s)} a_0^{(q_i,q_j,K; q_t)}
(\a^{(q_i,q_j)}(u)-1)(-q_0) + R_s(u=0).~~~\Label{WW-R-S}\eea
Finally,  the box coefficient is given by

\bea C[Q_i,Q_j]_{k=3} & = & {(K^2)^{2+n}\over 2} \left(
{\prod_{j=1}^{n+3} \gb{P_{(Q_j, Q_i);1}|\W{\W R}_j(u)|P_{(Q_j,
Q_i);2}}\over \gb{P_{(Q_j, Q_i);1}|K|P_{(Q_j,
Q_i);2}}^{n+2}\gb{P_{(Q_j, Q_i);1}|{\W Q}_t(u)|P_{(Q_j, Q_i);2}}}\right.
\nonumber \\
& & \left.- \prod_{s=1}^{n+3} \b_s^{(q_i,q_j,q_t;p_s)}{\gb{P_{(Q_j,
Q_i);1}|K|P_{(Q_j, Q_i);2}}\over \gb{P_{(Q_j, Q_i);1}|{\W
Q}_t(u)|P_{(Q_j, Q_i);2}}}+\{ P_{(Q_j, Q_i);1}\leftrightarrow
P_{(Q_j, Q_i);2}\}\right).~~\Label{k=3-final} \eea
%

%%%%%%%%%%%%%%%%%
%\subsubsection{Another point of view}
%%%%%%%%%%%%%%%%%

As an alternative to the expansion (\ref{ps-exp-k=3}) of $p_s$ in the  basis $q_i,q_j,q_t$, we consider another expansion:
\bea P_s = \sum_{\a=K, K_i, K_j, K_t} a_\a^{(K,K_i,K_j,K_t;P_s)}
K_\a.
~~~\Label{Ps-exp-k=3} \eea
By projecting equation (\ref{Ps-exp-k=3}) onto the vectorspace orthogonal to $K$, and comparing with (\ref{ps-exp-k=3}), we see that
\bea a_{w}^{(K,K_i,K_j,K_t;P_s)}=
a_w^{(q_i,q_j,q_t;p_s)},~~~~w=i,j,t.
\eea
%
%Via equation (\ref{Ps-exp-k=3}), we can relate $R_s(u)$ to the vectors $Q_\a(u)$, and derive the following expansion:
%%\footnote{It is important to notice that in following rewriting
%%we have used special form of $\b_s={-P_s\cdot K\over K^2}$.}
%%
%\bean R_s(u) & = & -\sqrt{1-u} P_s+{ -z(2P_s\cdot K)\over K^2} K \\& = &
%\sum_{\a= K_i, K_j, K_t} a_\a^{(K,K_i,K_j,K_t;P_s)} Q_\a(u)-\sum_{\a=K,
%K_i, K_j, K_t} a_\a^{(K,K_i,K_j,K_t;P_s)} {K_\a^2\over K^2} K\eean
%%
%From this we have
%%
%\bea \gb{P_{ji,1}|R_s|P_{ji,2}} & = &
%a_{K_t}^{(K,K_i,K_j,K_t;P_s)}\gb{P_{ji,1}| Q_t|P_{ji,2}}-\sum_{\a=K,
%K_i, K_j, K_t} a_\a^{(K,K_i,K_j,K_t;P_s)} {K_\a^2\over
%K^2}\gb{P_{ji,1}|K|P_{ji,2}}~~~\Label{pen-sep-1}\eea
%%

The advantage of the expansion (\ref{Ps-exp-k=3}) in four vectors is that we can solve for the coefficients explicitly, as in  (\ref{4D-exp}), and find
\bea a_t^{(q_i,q_j,q_t;p_s)} &=
&a_{t}^{(K,K_i,K_j,K_t;P_s)}={\eps(K_i,K_j, K, P_s)\over
\eps(K_i,K_j, K, K_t)}. ~~~~\Label{at-1}
\eea
Thus, using (\ref{big-beta-k-3}), we have
\bean
 \b_s^{(q_i,q_j,q_t;p_s)}
&=& -{K_i^2 \eps(P_s,K_j,K, K_t)+K_j^2 \eps(K_i,P_s,K, K_t) +K^2
\eps(K_i,K_j,P_s, K_t)+K_t^2 \eps(K_i,K_j,K, P_s)\over K^2
\eps(K_i,K_j,K, K_t)}.
\eean
With the  formulas  (\ref{at-1}) and (\ref{a-0-k=2}) for the expansion coefficients written in terms of input vectors, we can
simplify the following coefficient in the definition (\ref{wwr-k3}) of $\W{\W R}_s(u)$:
\bean a_t^{(q_i,q_j,q_t; p_s)} a_0^{(q_i,q_j,K; q_t)} & = &
-{\eps(K_i,K_j, K, P_s)\over \eps(K_t,K_i,K_j, K)}{\eps(K_t,
K_i,K_j,K)\over K^2 (q_0^{(q_i,q_j,K)})^2} ={\eps(P_s,K_i,K_j,
K)\over K^2 (q_0^{(q_i,q_j,K)})^2}=a_0^{(q_i,q_j,K; p_s)}.
\eean
We conclude that
\bea \W {\W R}_{s}(u)=\W R_s(u),~~~\Label{WWR-WR} \eea
where the definition of $\W R_s(u)$ is taken from (\ref{W-R-S-2}).

%%%%%%%%%%%%%
\subsubsection{Summary of $k=3$}
%%%%%%%%%%%%%

The pentagon coefficient is given by
\bea C[Q_i, Q_j, Q_t] = (K^2)^{3+n} \prod_{s=1}^{n+3}
\b_s^{(q_i,q_j,q_t;p_s)}.
~~~\Label{Pen-coeff-final}\eea
There are three boxes associated with the triplet of vectors $Q_i,Q_j,Q_t$.  We give the formula for the box involving $Q_i$ and $Q_j$; the other two may be obtained by exchanging indices.
\bea C[Q_i,Q_j]_{k=3} & = & {(K^2)^{2+n}\over 2} \left(
{\prod_{r=1}^{n+3} \gb{P_{(Q_j, Q_i);1}|{\W R}_r(u)|P_{(Q_j,
Q_i);2}}\over \gb{P_{(Q_j, Q_i);1}|K|P_{(Q_j,
Q_i);2}}^{n+2}\gb{P_{(Q_j, Q_i);1}|{\W Q}_t(u)|P_{(Q_j,
Q_i);2}}}\right.
\nonumber \\
& & \left.- \prod_{s=1}^{n+3} \b_s^{(q_i,q_j,q_t;p_s)}{\gb{P_{(Q_j,
Q_i);1}|K|P_{(Q_j, Q_i);2}}\over \gb{P_{(Q_j, Q_i);1}|{\W
Q}_t(u)|P_{(Q_j, Q_i);2}}}+\{ P_{(Q_j, Q_i);1}\leftrightarrow
P_{(Q_j, Q_i);2}\}\right),
~~\Label{k=3-final-1} \eea
where we have made the following definitions:
\bea  {\W R}_r(u) &=& {p_r\cdot q_0^{(q_i,q_j,K)}\over
(q_0^{(q_i,q_j,K)})^2}
(\a^{(q_i,q_j)}(u)-1)(-q_0^{(q_i,q_j,K)}) + R_r(u=0)~~~\Label{W-R-S-1}\\
\W Q_t(u) &= &{q_t\cdot q_0^{(q_i,q_j,K)}\over
(q_0^{(q_i,q_j,K)})^2}
(\a^{(q_i,q_j)}(u)-1)(-q_0^{(q_i,q_j,K)})+Q_t(u=0),~~~\Label{W-Q-t-1}\eea
and
\bea & & \b_s^{(q_i,q_j,q_t;p_s)} = \nonumber \\
& &  -{K_i^2 \eps(P_s,K_j,K, K_t)+K_j^2
\eps(K_i,P_s,K, K_t) +K^2 \eps(K_i,K_j,P_s, K_t)+K_t^2
\eps(K_i,K_j,K, P_s)\over K^2 \eps(K_i,K_j,K,
K_t)}. ~~~\Label{bs-k=3}\eea
%

%%%%%%%%%%%%%%%%%%%%%%%
\subsection{The case $k\geq 4$}
%%%%%%%%%%%%%%%%%%%%%%%%

If there at least four vectors $K_i$, then we can use four of them as a basis to expand the momentum vector $K$:
\bea K= \sum_{i=1}^4 a_i^{(1,2,3,4)} K_i.
 ~~~~\Label{K-exp-k=4}\eea
Using the expression (\ref{Q-K-def}) for $Q_i(u)$,
we find
\bean \sum_{i=1}^4 a_i^{(1,2,3,4)} Q_i(u)
%& = & -\sqrt{1-u} K +
%{\sum_{i=1}^4 a_i^{(1,2,3,4)} K_i^2 -2 z K^2\over K^2} K  \\
& = & {\sum_{i=1}^4 a_i^{(1,2,3,4)} K_i^2 -K^2\over K^2} K.
\eean

Use (\ref{P-RS}) to define the two null momenta $P_{i}(u)=Q_2(u)+ x_i Q_1(u)$.
Then, we have
\bean  {a_4^{(1,2,3,4)}\over \gb{P_1(u)|Q_3(u)|P_2(u)}}+
{a_3^{(1,2,3,4)}\over \gb{P_1(u)|Q_4(u)|P_2(u)}} &=& {\gb{P_1(u)| a_4^{(1,2,3,4)}
Q_4(u)+ a_3^{(1,2,3,4)} Q_3(u)|P_2(u)}\over
\gb{P_1(u)|Q_3(u)|P_2(u)} \gb{P_1(u)|Q_4(u)|P_2(u)}} \\
& = & {\gb{P_1(u)|\sum_{i=1}^4 a_i^{(1,2,3,4)} Q_i(u) |P_2(u)}\over
\gb{P_1(u)|Q_3(u)|P_2(u)} \gb{P_1(u)|Q_4(u)|P_2(u)}} \\
&=& {\sum_{i=1}^4 a_i^{(1,2,3,4)}
K_i^2 -K^2\over K^2} {\gb{P_1(u)|K |P_2(u)}\over \gb{P_1(u)|Q_3(u)|P_2(u)}
\gb{P_1(u)|Q_4(u)|P_2(u)}}.
 \eean
Therefore, we can derive the following identity:
\bea & & {1\over \gb{P_1(u)|Q_3(u)|P_2(u)} \gb{P_1(u)|Q_4(u)|P_2(u)}}  =  ~~~\Label{b-reduction-1}\\
& &  {K^2\over
\sum_{i=1}^4 a_i^{(1,2,3,4)} K_i^2 -K^2} {1\over \gb{P_1(u)|K|P_2(u)}}
\left({a_4^{(1,2,3,4)}\over \gb{P_1(u)|Q_3(u)|P_2(u)}}+ {a_3^{(1,2,3,4)}\over
\gb{P_1(u)|Q_4(u)|P_2(u)}} \right). \nonumber
\eea
Generalizing to our case with $k \geq 4$, we have
\bea & & {1\over \prod_{t=1,t\neq i,j}^k\gb{P_{ji,1}(u)|Q_t(u)|P_{ji,2}(u)}}
= ~~~\Label{b-reduction-2} \\
& & \sum_{t=1,t\neq i,j}^k {1\over
\gb{P_{ji,1}(u)|K|P_{ji,2}(u)}^{k-3}\gb{P_{ji,1}(u)|Q_t(u)|P_{ji,2}(u)}}\prod_{s=1,s\neq
i,j,t}^k \left( { a_s^{(i,j,t,s)}K^2\over  \sum_{\a=i,j,s,t}
a_\a^{(i,j,t,s)} K_\a^2 -K^2}\right). \nonumber
\eea
Thus, we can write
\bea & & {\prod_{r=1}^{k+n} \gb{P_{ji,1}(u)|R_r(u)
|P_{ji,2}(u)}\over \gb{P_{ji,1}(u)|K |P_{ji,2}(u)}^{n+2}
\prod_{t=1,t\neq i,j}^k\gb{P_{ji,1}(u)|Q_t(u) |P_{ji,2}(u)}}~~~\Label{k=4-exp-1}\\
& = & \sum_{t=1,t\neq i,j}^k \prod_{s=1,s\neq i,j,t}^k \left( {
a_s^{(i,j,t,s)}K^2\over  \sum_{\a=i,j,s,t} a_\a^{(i,j,t,s)} K_\a^2
-K^2}\right){\prod_{r=1}^{k+n} \gb{P_{ji,1}(u)|R_r(u)
|P_{ji,2}(u)}\over \gb{P_{ji,1}(u)|K |P_{ji,2}(u)}^{n+k-1}
\gb{P_{ji,1}(u)|Q_t(u) |P_{ji,2}(u)}}.
\nonumber
\eea
By this formula, we reduce the case with a given $n$, and $k \geq 4$, to the case with $k_{\rm eff}=3$, and $n_{\rm eff}=n+k-3$.
Using the results of the cases $k=2$ and $k=3$,  we  see
explicitly that indeed the coefficients of boxes are polynomials in
$u$.

The coefficients in (\ref{k=4-exp-1}) can be given more explicitly
using (\ref{a-sol}).  We denote the quantity in parentheses by
$1/\gamma_s^{(K_i,K_j;K_s,K_t)}$, and
\bea & & \gamma_s^{(K_i,K_j;K_s,K_t)}\equiv {
 \sum_{\a=i,j,s,t} a_\a^{(i,j,t,s)} K_\a^2
-K^2 \over a_s^{(i,j,t,s)}K^2 }
~~~~\Label{gamma-def}\\
& = & {
K_i^2\eps(K,K_j,K_s,K_t)+K_j^2\eps(K_i,K,K_s,K_t)+K_s^2\eps(K_i,K_j,K,K_t)
+K_t^2\eps(K_i,K_j,K_s,K)-K^2\eps(K_i,K_j,K_s,K_t)
\over K^2 \eps(K_i,K_j,K,K_t) }.
\nonumber \eea
The numerator of (\ref{gamma-def}) is symmetric in $K_i,K_j,K_s,K_t$; the denominator
singles out $K_s$, which is why we use the subscript $s$.

%%%%%%%%%%%%%%%%%%%%
\subsubsection{The total box coefficient}
%%%%%%%%%%%%%%%%%%%%%

We have just shown that when $k\geq 4$, we can use (\ref{k=4-exp-1}) to
reduce to terms with $k=3$, read out box coefficients for each
of these terms, and eventually add them all up.
This approach was useful to prove the polynomial property.  For computing amplitudes, we would like to carry out the summation once and for all.

We have used (\ref{b-reduction-1}) and
(\ref{b-reduction-2}) to derive (\ref{k=4-exp-1}). In each term on the right-hand side of
(\ref{k=4-exp-1}), there is one pentagon coefficient and one box
coefficient. The pentagon coefficients are uniquely associated to different pentagons, with the various $Q_t$'s along with $Q_i$ and $Q_j$, but the box
coefficients  all contribute to the same box, with only $Q_i$ and $Q_j$, so we must add them up.

Upon inspecting the final expression (\ref{k=3-final-1}) for box coefficients in the case $k=3$, we see that our task is to check that
equation (\ref{b-reduction-1}) still holds if we replace $Q_t(u)$ by $\W Q_t(u)$ and $P_i(u)$ by $P_i \equiv P_i(u=0)$.  Let us try to copy the derivation.
\bea & & {a_4^{(1,2,3,4)}\over \gb{P_1|\W Q_3(u)|P_2}}+
{a_3^{(1,2,3,4)}\over \gb{P_1|\W Q_4(u)|P_2}}
=  {\gb{P_1|
a_4^{(1,2,3,4)}\W Q_4(u)+ a_3^{(1,2,3,4)}\W Q_3(u)|P_2}\over
\gb{P_1|\W Q_3(u)|P_2} \gb{P_1|\W Q_4(u)|P_2}} \nonumber \\
& = & {\gb{P_1|\sum_{i=3}^4 a_{i}^{(1,2,3,4)} Q_i(u=0)
+(\a^{(q_i,q_j)}(u)-1)(-q_0)\sum_{i=3}^4 a_{i}^{(1,2,3,4)} {q_i\cdot
q_0\over q_0^2}|P_2}\over \gb{P_1|\W Q_3(u)|P_2} \gb{P_1|\W Q_4(u)|P_2}}.
~~~\Label{sec-split-k4}
\eea
Then second term in the numerator of (\ref{sec-split-k4}) is zero because $q_i\cdot q_0=q_j\cdot
q_0=K\cdot q_0=0$.  Further, we can extend the sum in the numerator of the first term to include $i=1,2$, since these terms are individually zero when contracted between spinors for $P_1$ and $P_2$.

Because
\bean \sum_{i=1}^4 a_{i}^{(1,2,3,4)} Q_i(u=0) =  \sum_{i=1}^{4}
a_{i}^{(1,2,3,4)} \a_i K,
\eean
 we have
\bean {a_4^{(1,2,3,4)}\over \gb{P_1|\W Q_3(u)|P_2}}+
{a_3^{(1,2,3,4)}\over \gb{P_1|\W Q_4(u)|P_2}}& =& {\gb{P_1|\sum_{i=1}^4
a_{i}^{(1,2,3,4)} \a_i K |P_2}\over \gb{P_1|\W Q_3(u)|P_2} \gb{P_1|\W
Q_4(u)|P_2}}.  \eean
It is now clear that  (\ref{b-reduction-1}) and
(\ref{b-reduction-2}) will still hold, if we replace $P_{ji}(u)\to P_{ji}(u=0)$
and $Q_t(u)\to \W Q_t(u)$.

Now we perform the summation, using (\ref{k=3-final-1}). There are two
terms. The first terms are collected to give
\bea{\prod_{r=1}^{k+n} \gb{P_{ji,1}(u=0)| {\W R}_r(u)
|P_{ji,2}(u=0)}\over \gb{P_{ji,1}(u=0)|K |P_{ji,2}(u=0)}^{n+2}
\prod_{t=1,t\neq i,j}^k\gb{P_{ji,1}(u=0)|\W Q_t(u)
|P_{ji,2}(u=0)}}.
\eea
The total from the second term from (\ref{k=3-final-1})
cannot be simplified further, since each
$\b_s^{(q_i,q_j,q_t;p_s)}$ depends on both $P_s$ and $K_t$.

%%%%%%%%%%%%%%%%%
\subsubsection{Results for $k\geq 4$}
%%%%%%%%%%%%%%%%%%%%%

The box coefficients are given by
\bea C[Q_i,Q_j]_{k\geq 4} & = & {(K^2)^{2+n}\over
2}\left\{{\prod_{r=1}^{k+n} \gb{P_{(Q_j,Q_i);1}| {\W R}_r(u)
|P_{(Q_j,Q_i);2}}\over \gb{P_{(Q_j,Q_i);1}|K |P_{(Q_j,Q_i);2}}^{n+2}
\prod_{t=1,t\neq
i,j}^k\gb{P_{(Q_j,Q_i);1}|\W Q_t(u) |P_{(Q_j,Q_i);2}}}\right.\nonumber  \\
& & \left.-\sum_{t=1,t\neq i,j}^k
{ \prod_{s=1}^{n+k} \b_s^{(q_i,q_j,q_t;p_s)} \over \prod_{w=1,w\neq i,j,t}^k
\gamma_w^{(K_i,K_j;K_w,K_t)} }
{
\gb{P_{(Q_j,Q_i);1}|K|P_{(Q_j,Q_i);2}}\over \gb{P_{(Q_j,Q_i);1}|\W
Q_t(u)|P_{(Q_j,Q_i);2}} }\right\}\nonumber \\ & & +\{
P_{(Q_j,Q_i);1}\leftrightarrow
P_{(Q_j,Q_i);2}\},
~~~\Label{k=4-box-equiv}\eea
where $\W R_r(u), \W Q_t(u),
\b_s^{(q_i,q_j,q_t;p_s)},\gamma_w^{(K_i,K_j;K_w,K_t)}$ are defined
in (\ref{W-R-S-1}), (\ref{W-Q-t-1}), (\ref{bs-k=3}) and
(\ref{gamma-def}), respectively. All $u$-dependence is inside $\W
R(u)$ and $\W Q(u)$. This form makes it easier to take the
derivative in (\ref{c-n-exp}).

Pentagon coefficients are given by
\bea C[Q_i, Q_j, Q_t] = (K^2)^{3+n} {\prod_{s=1}^{n+k}
\b_s^{(q_i,q_j,q_t;p_s)} \over \prod_{w=1,w\neq i,j,t}^k
\gamma_w^{(K_i,K_j;K_w,K_t)}
 }.
~~~\Label{Pen-coeff-final-2}\eea
%

%%%%%%%%%%%%%%%%%
\subsection{The degree of the polynomial}
%%%%%%%%%%%%%%%%%

In the cases $k=2$ and $k=3$, we have seen explicitly that the maximum degree of the polynomial is $[(n+2)/2]$.  For $k \geq 4$, the logic discussed after (\ref{k=4-exp-1}) implies only that the degree is no greater than $[(n+k-1)/2]$.

%Now let us discuss the degree of $u$ for the box coefficient. From
%the above discussion we can see that our procedure is that we reduce
%$k\geq 4$ to $k=3$ and finally the case of $k=2$. The degree of
%$k=2$ is given by $[m/2]$ where $m$ is the total number of momenta
%in numerator. By this logic, it seems that the degree of $u$ is
%given by $[m/2]$ for general $k$.

However, we can make a stronger claim by performing a different reduction.  In the previous subsection, we chose a reduction with a symmetric treatment of the factors $(-2\W \ell\cdot P_i)$ in the numerator of (\ref{I-inte}).  Alternatively, we could choose not to respect this symmetry in reducing the cases with $k \geq 3$.  For example, we can expand $P_s$ in terms of $K_i,K_j, K_t, K$ and use
\bea \gb{P_{ji,1}(u)|R_s(u)|P_{ji,2}(u)} & = & a_t^{(q_i,q_j,q_t;
p_1)} \gb{P_{ji,1}(u)|Q_t(u)|P_{ji,2}(u)} +\b_s^{(q_i,q_j,q_t; p_s)}
\gb{P_{ji,1}(u)|K|P_{ji,2}(u)}. \eea

This reduction translates to the following relation:
\bea \textrm{Box}[m,k,n] \rightarrow \textrm{Box}[m-1,k-1,n] +
\textrm{Box}[m-1,k,n-1]. \eea
Upon iteration, we arrive at $k=2$ for a fixed $n$.  Then, the degree of the polynomial is  $[(m-k+2)/2]=[(n+2)/2]$, which is what we wanted to show. (It is clear that the second term in
the above reduction will not have a higher degree than the first
term.)

%%%%%%%%%%%%%%%%%%%%%%
\section{Gluon example: $A(1^+,2^+,3^+,4^+,5^+)$}
%%%%%%%%%%%%%%%%%%%%%%

We now present an application to the five-gluon one-loop amplitude in Yang-Mills theory, which was first presented to all orders in $\eps$ in \cite{Bern:1995db}.  By supersymmetry arguments, the computation is equivalent to one with a scalar field circulating in the loop.

This configuration is totally symmetric, so we need only consider
any single, representative cut, say $C_{12}$. The others can be obtained by permuting labels.
The cut integrand within (\ref{I-inte}) is the product of two tree amplitudes given in \cite{Bern:1996ja}, with a factor of 2 for the two internal helicity choices:
\bean I_{12} & = & 2 A_L(-\ell_2, 1^+,2^+,-\ell_1) A_R
(\ell_1,3^+,4^+,5^+,\ell_2) \\ & = & 2 {\mu^2 [1~2]\over
\vev{1~2}((\ell_1-k_2)^2-\mu^2)} {- \mu^2 [5|k_{34} \ell_1|3]\over
\vev{3~4}\vev{4~5} ((\ell_1+k_3)^2-\mu^2) ((\ell_2+k_5)^2-\mu^2)} \\
& = & -2 {(\mu^2)^2 [1~2]\over \vev{1~2}\vev{3~4}\vev{4~5}} {
[5|k_{34} \W\ell|3]\over ((\W\ell-k_2)^2-\mu^2)
((\W\ell+k_3)^2-\mu^2) ((\W\ell+k_{34})^2-\mu^2)}
\\
 & = & -2 {(\mu^2)^2 [1~2]\over
\vev{1~2}\vev{3~4}\vev{4~5}} { (-2 \W\ell \cdot P_1) \over
((\W\ell-K_1)^2-\mu^2) ((\W\ell-K_2)^2-\mu^2)
((\W\ell-K_3)^2-\mu^2)}.
 \eean
where we have defined
\bea \W\ell = \ell_1 = k_{12}-\ell_2, \eea
and also that
\bean K = k_{12}, ~~~~ K_1 =
k_2,~~~~K_2=-k_3,~~~~K_3=-k_{34},~~~~P_1=[5~3]k_3+[5~4]\lambda_4\tilde\lambda_3.
\eean
According to our definitions in (\ref{I-inte}) and (\ref{mkn}), we
have
\bea
 k=3, \qquad m=1, \qquad n=-2.
\eea
Since triangles appear only when $n\geq-1$ and bubbles appear only
when $n\geq0$, we are left with only pentagon and box terms in this
case. Because $m=1$, the  degree of the polynomial in $u$ is 0.
Thus, we can set $u=0$ in our formula from the beginning.

From our formula (\ref{Pen-coeff-final}), we find that the pentagon coefficient is
\bea C_{pen} = -{2 (\mu^2)^2 [1~2] s_{12} \over
\vev{1~2}\vev{3~4}\vev{4~5}} \b_1^{(q_1,q_2,q_3;p_1)}. \eea
This coefficient is proportional to $(\mu^2)^2$.  The integral
 $I_5^D[(\mu^2)^2]$ is
$\cal O(\eps)$, which is what we expect of pentagons.

There are three boxes involved in this cut. Using the formula (\ref{k=3-final-1}), we find that
the box coefficient associated with $Q_1, Q_2$ is
\bea C[Q_1,Q_2] & = & {-(\mu^2)^2 [1~2]\over
\vev{1~2}\vev{3~4}\vev{4~5}} \left\{{\gb{P_{(Q_2,Q_1);1}| R_1
|P_{(Q_2,Q_1);2}}\over \gb{P_{(Q_2,Q_1);1}|Q_3 |P_{(Q_2,Q_1);2}}}-
\b_1^{(q_1,q_2,q_3;p_1)} { \gb{P_{(Q_2,Q_1);1}|K|P_{(Q_2,Q_1);2}}
\over
\gb{P_{(Q_2,Q_1);1}|Q_3|P_{(Q_2,Q_1);2}} }\right\}\nonumber \\
& & +\{ P_{(Q_2,Q_1);1}\leftrightarrow P_{(Q_2,Q_1);2}\} \nonumber
\\
& = & {-2(\mu^2)^2 [1~2]\over \vev{1~2}\vev{3~4}\vev{4~5}}
a_3^{(q_1,q_2,q_3; p_1)}, \eea
in which we have used (\ref{Rs-exp-k=3}) to simplify the expression.

Similar calculations give
\bea C[Q_1,Q_3] &=&  {-2(\mu^2)^2 [1~2]\over
\vev{1~2}\vev{3~4}\vev{4~5}}
a_3^{(q_1,q_3,q_2; p_1)}, \\
C[Q_2,Q_3] &=&  {-2(\mu^2)^2 [1~2]\over \vev{1~2}\vev{3~4}\vev{4~5}}
a_3^{(q_2,q_3,q_1; p_1)}. \eea

Now we check these coefficients against the result in the literature
\cite{Bern:1996ja, Brandhuber:2005jw}, which is (the factor ${i/
(4\pi)^{2-\eps}}$ is omitted, and we also changed the result to our
convention of the basis definition (\ref{DIN-def}))
\bea  & & A(1^+,2^+,3^+,4^+,5^+) =
\frac{\eps(1-\eps)}{\vev{12}\vev{23}\vev{34}\vev{45}\vev{51}} \Big(
- 4i(4-2\eps) ~\eps(k_1,k_2,k_3,k_4)I_5^{D+6}[1] +
s_{23}s_{34}I_4^{D+4,(1)}[1]  \nonumber\\
& & \qquad \qquad \qquad \qquad \quad \quad +
s_{34}s_{45}I_4^{D+4,(2)}[1] + s_{45}s_{51}I_4^{D+4,(3)}[1] +
s_{51}s_{12}I_4^{D+4,(4)}[1] + s_{12}s_{23}I_4^{D+4,(5)}[1]
\Big) \nonumber\\
& & \qquad \qquad \qquad  \quad \quad ~~~ =
\frac{1}{\vev{1~2}\vev{2~3}\vev{3~4}\vev{4~5}\vev{5~1}} \Big( 8 i
\eps(k_1,k_2,k_3,k_4) I_5^{D}[\mu^6]
-s_{23}s_{34}I_4^{D,(1)}[\mu^4] \nonumber\\
& & \qquad \qquad \qquad \qquad \quad \quad -
s_{34}s_{45}I_4^{D,(2)}[\mu^4] -s_{45}s_{51}I_4^{D,(3)}[\mu^4]
-s_{51}s_{12}I_4^{D,(4)}[\mu^4] -s_{12}s_{23}I_4^{D,(5)}[\mu^4]
\Big).
 \eea
We have used $I_4^D[\mu^4]=-\eps(1-\eps)I_4^{D+4}[1]$ and
$I_5^D[\mu^6]=-\eps(1-\eps)(2-\eps)I_5^{D+6}[1]$.
In Appendix \ref{dsb}, we discuss various recursive relations and dimensional shift identities.  We now apply the identity (\ref{gen-rec-mu}) to get
\bea I_5^D[\mu^6]=\left(-\frac{1}{\Delta_5}\right) I_5^D[\mu^4]
+{1\over2}\sum_{i=1}^5 \left(-\frac{\gamma_{5,i}}{\Delta_5}\right)~
I_4^{D,(i)}[\mu^4] . \eea
Then we see that we should be able to reproduce the following correspondences.  For $C[Q_1,Q_2]$,
\bea {2 [1~2]\over \vev{1~2}\vev{3~4}\vev{4~5}} a_3^{(q_1,q_2,q_3;
p_1)} = \frac{1}{\vev{1~2}\vev{2~3}\vev{3~4}\vev{4~5}\vev{5~1}}
\left(s_{12}s_{23} + 4i\eps(k_1,k_2,k_3,k_4)
\frac{\gamma_{5,5}}{\Delta_5}\right); \eea
for the pentagon,
\bea {2 [1~2] s_{12} \over \vev{1~2}\vev{3~4}\vev{4~5}}
\b_1^{(q_1,q_2,q_3;p_1)} =
\frac{1}{\vev{1~2}\vev{2~3}\vev{3~4}\vev{4~5}\vev{5~1}}
\left(8i\eps(k_1,k_2,k_3,k_4)\frac{1}{\Delta_5}\right) . \eea
We have checked that these equations (as well as the ones derived from the other two boxes) are consistent with our definitions.

%%%%%%%%%%%%%%%%%%%%%%%%%

%%%%%%%%%%%%%%%%%%%%%%
\section{Discussion}
%%%%%%%%%%%%%%%%%%%%%%

From the $u$-dependent formulas for 4-dimensional integral
coefficients given in \cite{Britto:2007tt}, we have now given
simpler versions, explicit proofs that they are polynomials, and the
formula (\ref{c-n-exp}) needed for the final evaluation in $D$
dimensions.  In this section, we make  some remarks comparing our
results with two recent papers, \cite{Ossola:2008xq} and
\cite{Giele:2008ve}.

The authors of \cite{Ossola:2008xq} discussed the calculation of
rational terms. The rational
contribution may be split into two parts (eq. (3)
of \cite{Ossola:2008xq}), namely a term depending on $\W q^2$ (which is $\mu^2$ in
our notation) and the 4-dimensional part. For the former term, the
authors of \cite{Ossola:2008xq} reduce the calculation into
effective Feynman diagrams. In our approach, we do not distinguish
these two terms; they are treated on the same footing by using
dimensionally shifted master integrals.
To deal with the 4-dimensional terms independent of $\W q^2$,
\cite{Ossola:2008xq}
proposed  the mass shifted method (eq. (16) of \cite{Ossola:2008xq})
and following expansion (eqs. (17), (18), (20) of
\cite{Ossola:2008xq}).
In our terminology, these are the coefficients of $u^a$, which we have
discussed. The proposal of \cite{Ossola:2008xq} is to choose different
values of $\W q^2$, while here we use
the derivative.  We
could also choose to substitute numerical values of $u$, and
then find the coefficient from a linear equation,
  as detailed recently in \cite{Britto:2008vq}.
In another recent paper \cite{Mastrolia:2008jb},  this same numerical approach is implemented
in the context of \cite{Ossola:2008xq}.

The authors of \cite{Giele:2008ve} treat $s_e^2$ (which is $u$ in
our notation) as an effective dimension.  Thus they are able to use
a 5-dimensional cut to read off the pentagon coefficient. To get the
coefficients, they work in two different dimensions, $D_1$ and
$D_2$. The paper \cite{Giele:2008ve} has thus given a way to deal
with the problem of the polarization tensor of a gluon or fermion in
arbitrary dimension $D$. By choosing appropriate loop momenta and
solving a linear system of equations, they can separate the
coefficients into spurious terms and the terms with various powers
of $s_e^2$. As in our approach, the terms with non-zero powers of
$s_e^2$ will contribute to the coefficients of dimensionally shifted
master integrals.

The methods of \cite{Giele:2008ve,Ossola:2008xq} have been
implemented numerically
\cite{Giele:2008ve,Mastrolia:2008jb,Binoth:2008kt,Giele:2008bc} and
have been shown to be stable and efficient.\footnote{Note added in
  revised version: There has recently appeared a numerical implementation of
  another technique for getting rational parts of one-loop amplitudes \cite{Badger:2008cm},
based on the generalized-unitarity formalism of \cite{Forde:2007mi,Berger:2008sj} combined with an expansion
in $\mu^2$.}  We have not yet attempted
a numerical implementation of the procedure given in this paper, and
we leave its assessment to future work.
 Analytically, our algebraic
expressions are  the most general, since we have not assumed
renormalizability, and the power of $q$ in the numerator can be
arbitrarily high.

%%%%%%%%%%%%%%%%%%%%%%%%%%%%%%%
\acknowledgments
%%%%%%%%%%%%%%%%%%%%%%%%%%%%%%%
We are grateful to  C.-J. Zhu for helpful discussions.
RB is supported by Stichting FOM. BF  is
 supported by Qiu-Shi Professor Fellowship from Zhejiang University,
 China.
GY is supported by funds from the National Natural Science Foundation of  China with
grant Nos. 10475104 and 10525522.

%%%%%%%%%%%%%%%%%%%%%%%%%
\appendix
%%%%%%%%%%%%%%%%%%%%%%

%%%%%%%%%%%%%%%%%%%%%%%%%%%%%%%%%%%%%%%%%%%%%%%%%%%%%%%%%%%%%%%%%%%%%
\section{\label{dsb}The scalar integrals and dimensional shift identities}
%%%%%%%%%%%%%%%%%%%%%%%%%%%%%%%%%%%%%%%%%%%%%%%%%%%%%%%%%%%%%%%%%%%%%
The $\frak D$-dimensional scalar integral  is defined to be
\bea I_n^{\frak D}[1] & \equiv & -i (4\pi)^{{\frak D}/2} \int \frac{d^{\frak
D}p}{(2\pi)^{\frak D}}
\frac{1}{p^2(p-K_1)^2(p-K_1-K_2)^2\cdots(p+K_{n})^2}. \eea
We will use $\frak D$ to denote the dimensionality, so that we can specifically set
\bea
D\equiv 4-2\eps.
\eea
 We also define a very useful symmetric matrix, $S$, as follows:
\bea S & \equiv &  - {1\over 2} \, \begin{pmatrix} 0 & K_1^2 &
(K_1+K_2)^2 & \cdots & (K_1+K_2+\cdots K_{n-1})^2 \cr
* & 0 & K_2^2 & \cdots & (K_2+K_3+\cdots K_{n-1})^2 \cr
\vdots & \vdots & \vdots & \vdots & \vdots \cr
* & * & * & 0  &     K_{n-1}^2 \cr
* & * & * & *  & 0
\end{pmatrix} = - {1\over 2} \,
\begin{pmatrix}
0 & s_1 & s_{12} & \cdots & s_n \cr
* & 0 & s_2 & \cdots & s_{n1} \cr
\vdots & \vdots & \vdots & \vdots & \vdots \cr
* & * & * & 0  &     s_{n-1} \cr
* & * & * & *  & 0
\end{pmatrix}. ~~~\Label{useful-S}\eea
%
%which is an $n\times n$ symmetric matrix of external kinematic
%variables, and we have defined
%%
%\begin{equation}
%s_{ij\cdots m} \equiv (K_i+K_j+\cdots+K_m)^2
%\end{equation}
%%

If there are explicit powers of $\mu^2$ in the numerator, we expand in $\eps$ as follows,
\bea I_n^D[(\mu^2)^k] = \frac{\Gamma(k-\eps)}{\Gamma(-\eps)}
I_n^{D+2k}[1] = -\eps~\Gamma(k)~I_n^{D+2k}[1]+ {\cal O}(\eps) \eea
and deduce that we only need to calculate the coefficient of the $1/\eps$ term of
$I_n^{D+2k}[1]$, in order to get the rational term. For bubbles and triangles,
we need to consider $k\geq1$; for boxes, we need to consider $k\geq2$;
and for pentagons, we need to consider $k\geq3$.

We will use two ways to deal with the higher-dimensional scalar
integral, mainly following \cite{IntegralRecursion}.

The first way is by calculating the integral directly, using Feynman
parametrization:
\bea I_n^{\frak D}[1] & = & (-1)^n\, \Gamma(n-{\frak D}/2) \,
\int_0^1 { d a_1 \cdots d a_n} \, { \delta( 1 - \sum_i a_i) \over (
a \cdot S \cdot a )^{n -{{\frak D}\over 2}} } ~, \eea
where
$$a \cdot S \cdot a  = \sum_{i,j=1}^n a_i\,a_j\, S_{ij} ~.$$
This integral is easy in the cases of bubbles and one-mass or two-mass triangles.   However, for three-mass triangles, boxes and pentagons, the integral is complicated.

The second way is by using a recursive relation, which reduces the
higher-dimensional scalar integrals to lower-dimensional and
lower-point scalar integrals \cite{IntegralRecursion}:
\bea I^{{\frak D}+2}_n[1] = {1\over (n-1-{\frak D})~ \Delta_n} \,
\left[ 2 \, I_n^{\frak D}[1] + \sum_{i=1}^n \gamma_{n,i} \,
I_{n-1}^{{\frak D},(i)}[1] \right], \label{higherDD} \eea
where
\bea \gamma_{n,i} =  \sum_{j=1}^n S_{ij}^{-1}, \qquad \Delta_n =
\sum_{i=1}^n \gamma_{n,i} . \eea
If ${\frak D}=D+2k-2$, then we have
\bea I^{D+2k}_n[1] & = & {1\over k-{n-3\over 2}-\eps} \, \left[
\left( -{1 \over \Delta_n} \right) \, I_n^{D+2(k-1)}[1] + {1\over2}
\sum_{i=1}^n \left( -{\gamma_{n,i} \over \Delta_n} \right) \,
I_{n-1}^{{D+2(k-1)},(i)}[1] \right] . \quad \Label{gen-rec} \eea
Similarly, we can write
\bea I_n^{D}\left[(\mu^2)^k\right] &=&
\frac{k-1-\eps}{k-{n-3\over2}-\eps} \left[ \left( -{1 \over
\Delta_n} \right) I_n^{D}\left[(\mu^2)^{k-1}\right]
+\frac{1}{2}\sum_{i=1}^n \left( -{\gamma_{n,i} \over \Delta_n}
\right)~ I_{n-1}^{D,(i)}\left[(\mu^2)^{k-1}\right]\right] . \quad
\Label{gen-rec-mu} \eea
The recursive relations are very convenient in dealing with three-mass
triangle and higher point cases.
For one-mass and two-mass triangles, the matrix $S$ is singular, so these recursive relations are not well defined.  However, it is possible to recover the results from massless limits of the three-mass triangle.  In practice, then, we can always use these recursive relations, taking a massless limit at the end in special cases (also boxes and pentagons).

In the following, we give compact recursive formulas for bubble,
triangle, box and pentagon, for convenient automated evaluation.

\subsection{Bubble}
For the bubble, the matrix $S$ defined in (\ref{useful-S}) becomes
\bea S=-{1\over2} \begin{pmatrix} 0 & K^2 \\ K^2 & 0
\end{pmatrix}, \quad S^{-1}=-2 \begin{pmatrix} 0 & \frac{1}{K^2} \\ \frac{1}{K^2} & 0
\end{pmatrix}, \eea
so
\bea \gamma_{2,1}=\gamma_{2,2}= -\frac{2}{K^2}~, \quad
\Delta_2=-\frac{4}{K^2}~, \eea
and
\bea a \cdot S \cdot a = -a_1 a_2 K^2 ~. \eea

Using the recursive relation (\ref{gen-rec}), we find
\bea I_2^{D+2k}[1] &=& \frac{1}{k+{1\over2}-\eps}
\left(-\frac{1}{\Delta_2}\right) I_2^{D+2(k-1)}[1] \nonumber\\ &=&
\frac{1}{(1+{1\over2})(2+{1\over2})\cdots(k+{1\over2})}
\left(\frac{K^2}{4}\right)^k I_2^D[1]+ {\cal O}(\eps) \nonumber\\
&=& \frac{\sqrt{\pi}/2}{\Gamma(k+\frac{3}{2})}
\left(\frac{K^2}{4}\right)^k \frac{1}{\eps}+ {\cal O}(\eps^0). \eea
Alternative, we can use the Feynman parametrization to calculate
it directly.
\bea I_2^{D+2k}[1] &=& \Gamma(-k+\eps)\int_0^1 d a_1 d a_2
~\delta(1-a_1-a_2) ~(-a_1 a_2 K^2)^{k-\eps} \nonumber\\
&=& \Gamma(-k+\eps)\frac{\Gamma(k+1-\eps)^2}
{\Gamma(2k+2-2\eps)}(-K^2)^{k-\eps} \nonumber\\ &=&
\frac{\Gamma(k+1) (K^2)^k}{\Gamma(2k+2)}~\frac{1}{\eps} + {\cal
O}(\eps^0). \eea
\subsection{Triangle}
The matrix $S$ is
\bea S = -{1\over2} \begin{pmatrix}0 & K_1^2 & K_3^2 \\ K_1^2 & 0 &
K_2^2 \\ K_3^2 & K_2^2 & 0 \end{pmatrix} = -{1\over2}
\begin{pmatrix}0 & s_1 & s_3 \\ s_1 & 0 & s_2
\\ s_3 & s_2 & 0 \end{pmatrix}.
\eea
Its inverse is
\bea \quad S^{-1}
= -\frac{1}{s_1s_2s_3} \begin{pmatrix} -s_2^2 & s_2s_3 & s_1s_2 \\
s_2s_3 & -s_3^2 & s_3s_1 \\ s_1s_2 & s_3s_1 & -s_1^2 \end{pmatrix},
\eea
so
\bea \gamma_{3,1} &=& \frac{s_2(s_2-s_3-s_1)}{s_3s_1}, \quad
\gamma_{3,2}=\frac{s_3(s_3-s_1-s_2)}{s_1s_2}, \quad
\gamma_{3,3}=\frac{s_1(s_1-s_2-s_3)}{s_2s_3}, \nonumber\\
\Delta_3 &=&
\frac{s_1^2+s_2^2+s_3^2-2(s_1s_2+s_2s_3+s_3s_1)}{s_1s_2s_3}, \eea
and
\bea a \cdot S \cdot a = -(a_1 a_2 K_1^2 + a_2 a_3 K_2^2 + a_3 a_1
K_3^2). \eea

One-mass and two-mass triangles can be evaluated by
Feynman parametrization. If $K_3^2=0$, then
\bea I_{3; 2m}^{D+2k}[1] & = & - \Gamma(-k+1+\eps) \int_0^1 d a_1 d
a_2 d a_3 ~\delta(1-a_1-a_2-a_3) ~(-a_1 a_2 K_1^2 - a_2 a_3
K_2^2)^{k-1-\eps} \nonumber\\ & = & -
\frac{\Gamma(k)}{\Gamma(2k+1)}~
\frac{(K_1^2)^k-(K_2^2)^k}{K_1^2-K_2^2}~\frac{1}{\eps} + {\cal
O}(\eps^0). \eea

If, in addition, $K_2^2=0$, then
\bea I_{3; 1m}^{D+2k}[1] = - \frac{\Gamma(k)}{\Gamma(2k+1)}~
(K_1^2)^{k-1}~\frac{1}{\eps} + {\cal O}(\eps^0). \eea

For the three-mass triangle, we use the recursive relation
(\ref{gen-rec}) repeatedly, and obtain
\bea I_3^{D+2k}[1] & = & \frac{1}{k-\eps} \left[
\left(-\frac{1}{\Delta_3}\right) I_3^{D+2(k-1)}[1] + \frac{1}{2}
\sum_{i=1}^3 \left(-\frac{\gamma_{3,i}}{\Delta_3}\right) ~
I_2^{D+2(k-1),(i)}[1]\right] \nonumber\\
& = & {1\over2}
\sum_{\ell=0}^{k-1}\frac{\Gamma(\ell+1)}{\Gamma(k+1)}
\left(-\frac{1}{\Delta_3}\right)^{k-\ell} \sum_{i=1}^3 \gamma_{3,i}~
I_2^{D+2\ell,(i)}[1] +  {\cal O}(\eps^0), \quad
\Label{triangle-rec}\eea
where we make use of our previously derived bubble result.
The first several cases, written explicitly, are
\bea I_3^{D+2}[1] &=& - \frac{1}{2 \eps} + {\cal O}(\eps^0) \nonumber\\
I_3^{D+4}[1] &=& - \frac{1}{24 \eps}(s_1+s_2+s_3) + {\cal O}(\eps^0) \nonumber\\
I_3^{D+6}[1] &=& - \frac{1}{360 \eps}
(s_1^2+s_2^2+s_3^2+s_1s_2+s_2s_3+s_3s_1) + {\cal O}(\eps^0) \nonumber\\
I_3^{D+8}[1] &=& - \frac{1}{6720 \eps}\Big(s_1^3+s_2^3+s_3^3
+s_1^2s_2+s_1s_2^2+s_2^2s_3+s_2s_3^2+s_3^2s_1+s_3s_1^2
+{4\over3}s_1s_2s_3\Big) + {\cal O}(\eps^0) \nonumber \eea
Note that  $\Delta_3$ in (\ref{triangle-rec}) has cancelled out of the numerator and denominator.
% The divergent terms are polynomials in $s_i$, and the power
%can be counted from the integral directly.

We have verified that the results for one-mass and
two-mass triangles are consistent with the massless limit of the three-mass triangle result.

\subsection{Box}

The matrix $S$ is
\bea S= -{1\over2} \begin{pmatrix}0 & K_1^2 & (K_1+K_2)^2 & K_4^2 \\
K_1^2 & 0 & K_2^2 & (K_2+K_3)^2 \\ (K_1+K_2)^2 & K_2^2 & 0 & K_3^2 \\
K_4^2 & (K_2+K_3)^2 & K_3^2 & 0
\end{pmatrix} = -{1\over2}
\begin{pmatrix}0 & s_1 & s_{12} & s_4 \\
s_1 & 0 & s_2 & s_{23} \\ s_{12} & s_2 & 0 & s_3 \\
s_4 & s_{23} & s_3 & 0
\end{pmatrix}. \eea

Using the recursive relation (\ref{gen-rec}) repeatedly, we find
\bea I_4^{D+2k}[1] & = & \frac{1}{k-{1\over2}-\eps} \left[
\left(\frac{-1}{\Delta_4}\right) I_4^{D+2(k-1)}[1] +\frac{1}{2}
\sum_{i=1}^4 \left(-\frac{\gamma_{4,i}}{\Delta_4}\right)~
I_{3}^{D+2(k-1),(i)}[1]\right] \nonumber\\ & = & {1\over2}
\sum_{\ell=1}^{k-1}\frac{\Gamma(\ell+{1\over2})}{\Gamma(k+{1\over2})}
\left(\frac{-1}{\Delta_4}\right)^{k-\ell} \sum_{i=1}^4 \gamma_{4,i}~
I_3^{D+2\ell,(i)}[1] +  {\cal O}(\eps^0). \quad \Label{box-rec}\eea
Here we use the identities for triangles.  The first few cases, listed explicitly, are
\bea I_4^{D+4}[1] &=&
\frac{1}{6\eps} + {\cal O}(\eps^0) \nonumber\\
I_4^{D+6}[1] &=& \frac{1}{120 \eps}(s_1+s_2+s_3+s_4+s_{12}+s_{23})
 + {\cal O}(\eps^0) \nonumber\\
I_4^{D+8}[1] &=& \frac{1}{2520 \eps}\Big(s_1^2+s_2^2+s_3 ^2 +s_4^2
+s_{12}^2+s_{23}^2 +s_1s_2+s_2s_3+s_3s_4+s_4s_1 \nonumber\\
&& ~~~~~~~~~+(s_{12}+s_{23})(s_1+s_2+s_3+s_4)
+{1\over2}(s_1s_3+s_2s_4+s_{12}s_{23})\Big) + {\cal O}(\eps^0)
\nonumber \eea
Notice that the factor $\Delta_4$ in (\ref{box-rec}) cancels out.

\subsection{Pentagon}
The matrix $S$ for the pentagon is
\bea S= -{1\over2}
\begin{pmatrix}0 & s_1 & s_{12} & s_{45} & s_5 \\
s_1 & 0 & s_2 & s_{23} & s_{51} \\ s_{12} & s_2 & 0 & s_3 & s_{34} \\
s_{45} & s_{23} & s_3 & 0 & s_4 \\ s_5 & s_{51} & s_{34} & s_4 & 0
\end{pmatrix}.
 \eea

Using the recursive relation (\ref{gen-rec}) repeatedly, we find
\bea I_5^{D+2k}[1] & = & \frac{1}{k-1-\eps} \left[
\left(\frac{-1}{\Delta_5}\right) I_5^{D+2(k-1)}[1]
 + \frac{1}{2}\sum_{i=1}^5
\left(\frac{-\gamma_{5,i}}{\Delta_5}\right)~
I_{5-1}^{D+2(k-1),(i)}[1]\right] \nonumber\\ & = & {1\over2}
\sum_{\ell=2}^{k-1}\frac{\Gamma(\ell)}{\Gamma(k)}
\left(\frac{-1}{\Delta_5}\right)^{k-\ell} \sum_{i=1}^5 \gamma_{5,i}~
I_4^{D+2\ell,(i)}[1] +  {\cal O}(\eps^0). \quad \Label{pen-rec}\eea
The first few identities we derive this way are
\bea I_5^{D+6}[1] & = &
- \frac{1}{24 \eps} + {\cal O}(\eps^0) \nonumber\\
I_5^{D+8}[1] & = & - \frac{1}{720 \eps}(s_1+s_2+s_3+s_4+s_5
+s_{12}+s_{23}+s_{34}+s_{45}+s_{51}) + {\cal O}(\eps^0) \nonumber\\
I_5^{D+10}[1] & = & - \frac{1}{20160
\eps}\Big(s_1^2+s_{12}^2+s_1s_2+s_{12}s_{34}
+s_1(s_{12}+s_{23}+s_{45}+s_{51})
+{1\over2}(s_1s_3+s_1s_{34}+s_{12}s_{23}) \nonumber\\ &&
~~~~~~~~~~~~~ + \textmd{four cyclic} \Big) + {\cal O}(\eps^0)
\nonumber \eea
The factor $\Delta_5$ in (\ref{pen-rec}) cancels out.

%%%%%%%%%%%%%%%%%
\section{\label{explicit}Explicit expressions for triangle coefficients}
%%%%%%%%%%%%%%%%%

In this appendix we collect some simplified expressions for triangle coefficients
with $n\leq 2$. This is sufficient in renormalizable theories. For
general $n$, we can always go back to the general formula (\ref{n=0-1}).

For $n=-1$:
\bea C[Q_s, K]_{n=-1} & = & {1\over 2}\left(
{\prod_{j=1}^{k-1}\gb{P_{(q_s,K);1}|P_j|P_{(q_s,K);2}}\over
\prod_{t=1,t\neq s}^k \gb{P_{(q_s,K);1}|K_t|P_{(q_s,K);2}}}+
\{P_{(q_s,K);1}\leftrightarrow P_{(q_s,K);2}\}\right).
\eea

For $n=0$:
\bea C[Q_s,K]_{n=0} & = & -{\a_s (K^2)^2\over
{\Delta(q_s,K)}}\left\{{\prod_{j=1}^{k} \gb{P_{(q_s,K);1}|P_j
|P_{(q_s,K);2}}\over \prod_{t=1,t\neq s}^k \gb{P_{(q_s,K);1}|K_t
|P_{(q_s,K);2}}}\left( -\sum_{j=1}^{k}{ 2\W p_j\cdot q_s\over
\gb{P_{(q_s,K);1}|P_j |P_{(q_s,K);2}}}\right.\right. \nonumber \\ &
& \left.\left.+\sum_{t=1, t\neq s}^k{ 2\W q_t\cdot q_s\over
\gb{P_{(q_s,K);1}|K_t |P_{(q_s,K);2}}}\right)+\{
P_{(q_s,K);1}\leftrightarrow P_{(q_s,K);2}\} \right\}.
\eea

For $n=1$, we have linear $u$-dependence:
\bea C[Q_s,K]_{n=1} & = & { (K^2)^4 \a_s^2 \over  \Delta(q_s,K)^2 }
\left( {\prod_{j=1}^{k+1}\gb{P_{(q_s,K);1} |P_j | P_{(q_s,K);2}}
\over \prod_{t=1,t\neq s}^k \gb{P_{(q_s,K);1} |K_t | P_{(q_s,K);2}}}
({\cal F}_1^2+ {\cal F}_2) +\{P_{(q_s,K);1}\leftrightarrow
P_{(q_s,K);2}\}\right),
\eea
where
\bea {\cal F}_1 & = & \left(-\sum_{j=1}^{k+1}{ (2\W p_j\cdot
q_s)\over \gb{P_{(q_s,K);1}|P_j|P_{(q_s,K);2}}}+\sum_{t=1, t\neq s}{
(2\W q_t\cdot q_s)\over \gb{P_{(q_s,K);1}|K_t |P_{(q_s,K);2}}}\right) \\
{\cal F}_2 & = & -\sum_{j=1}^{k+1}\left({ (2\W p_j\cdot q_s)\over
\gb{P_{(q_s,K);1}| P_j |P_{(q_s,K);2}}}\right)^2+\sum_{j=1}^{k+1}
{1\over 2}\left( (1-u){q_s^2\over
\a_s^2K^2}+1\right){\gb{P_{(q_s,K);2}|P_j|P_{(q_s,K);1}}
\over \gb{P_{(q_s,K);1}|P_j |P_{(q_s,K);2}}} \nonumber \\
& &  +\sum_{t=1, t\neq s}^k\left({(2\W q_t\cdot q_s)\over
\gb{P_{(q_s,K);1}|K_t |P_{(q_s,K);2}}}\right)^2-\sum_{t=1, t\neq
s}^k {1\over 2} \left( (1-u){q_s^2\over
\a_s^2K^2}+1\right){\gb{P_{(q_s,K);2}|K_t|P_{(q_s,K);1}}\over
\gb{P_{(q_s,K);1}|K_t|P_{(q_s,K);2}}}\eea

Finally, for $n=2$, we again have linear $u$-dependence:
\bea C[Q_s,K]_{n=2} & = &-{2\over 3} {(K^2)^{6}\over
{\Delta(q_s,K)}^{3}} \left( {\prod_{j=1}^{k+2}\gb{P_{(q_s,K);1} |P_j
| P_{(q_s,K);2}} \over \prod_{t=1,t\neq s}^k \gb{P_{(q_s,K);1} |K_t
| P_{(q_s,K);2}}} ({\cal F}_1^3+ 3{\cal F}_1 {\cal F}_2 +{\cal F}_3)
+\{P_{(q_s,K);1}\leftrightarrow P_{(q_s,K);2}\}\right), \nonumber \eea
where
\bea {\cal F}_1 & = & \left(-\sum_{j=1}^{k+2}{ (2\W p_j\cdot
q_s)\over \gb{P_{(q_s,K);1}|P_j|P_{(q_s,K);2}}}+\sum_{t=1, t\neq
s}^k{ (2\W q_t\cdot q_s)\over \gb{P_{(q_s,K);1}|K_t |P_{(q_s,K);2}}}\right), \\
{\cal F}_2 & = & -\sum_{j=1}^{k+2}\left({ (2\W p_j\cdot q_s)\over
\gb{P_{(q_s,K);1}| P_j |P_{(q_s,K);2}}}\right)^2+\sum_{j=1}^{k+2}
{1\over 2}\left( (1-u){q_s^2\over
\a_s^2K^2}+1\right){\gb{P_{(q_s,K);2}|P_j|P_{(q_s,K);1}}
\over \vev{P_{(q_s,K);1}|P_j |P_{(q_s,K);2}}} \nonumber \\
& &  +\sum_{t=1, t\neq s}^k\left({(2\W q_t\cdot q_s)\over
\gb{P_{(q_s,K);1}|K_t |P_{(q_s,K);2}}}\right)^2-\sum_{t=1, t\neq
s}^k {1\over 2} \left( (1-u){q_s^2\over
\a_s^2K^2}+1\right){\gb{P_{(q_s,K);2}|K_t|P_{(q_s,K);1}}\over
\gb{P_{(q_s,K);1}|K_t|P_{(q_s,K);2}}},
\eea
and
\bea {\cal F}_3 & = & \left\{ -\sum_{j=1}^{k+2}2\left({ (2\W
p_j\cdot q_s)\over \gb{P_{(q_s,K);1}|P_j |P_{(q_s,K);2}}}\right)^3+
\sum_{t=1, t\neq s}^k 2\left({ (2\W q_t\cdot q_s)\over
\gb{P_{(q_s,K);1}|K_t |P_{(q_s,K);2}}}\right)^3 \right. \nonumber\\
& & +\sum_{j=1}^{k+2} {3\over 2}\left( (1-u){q_s^2\over
\a_s^2K^2}+1\right){\gb{P_{(q_s,K);2}|P_j|P_{(q_s,K);1}} (2\W
p_j\cdot q_s)\over
\gb{P_{(q_s,K);1}|P_j|P_{(q_s,K);2}}^2} \nonumber \\
& & \left. -\sum_{t=1, t\neq s}^k {3\over 2} \left( (1-u){q_s^2\over
\a_s^2K^2}+1\right){\gb{P_{(q_s,K);2}|K_t|P_{(q_s,K);1}} (2\W
q_t\cdot q_s)\over
\gb{P_{(q_s,K);1}|Q_t|P_{(q_s,K);2}}^2}\right\}.
\eea
%

%%%%%%%%%%%%%%%%%%%%
\section{\label{bubbleproof}Proof of the polynomial property of bubble coefficients}
%%%%%%%%%%%%%%%%%%%%

Here we present a proof that the bubble coefficients are polynomials in $u$.
For this proof, we make use of their derivation from spinor integrals \cite{Britto:2007tt}, along with certain results of Ossola, Papadopoulos and Pittau (OPP) \cite{Ossola:2006us} to analyze the integrand.

Given the cut integral (\ref{I-inte}), bubble coefficients are given
by the sum of residues at the poles of the following function (see
Appendix B of \cite{Britto:2007tt}). \footnote{In this discussion,
we drop prefactors independent of loop momentum, as well as the
possible prefactor $c(\mu^2)$. Also, we have used $(K-s\eta)$
instead of $(K+s\eta)$ in (\ref{bub-exp}).
This change is compensated by dropping the factor $(-1)^q$.}
\bea \sum_{q=0}^n \left. {1\over q!} {d^q B_{n,n-q}(s)\over
ds^q}\right|_{s=0},~~~~\Label{gen-n}\eea
where the residues of $B_{n,t}(s)$ are taken before the derivative in $s$, and the function $B_{n,t}(s)$ is defined to be
\bea B_{n,t}(s) \equiv {\gb{\ell|\eta|\ell}^t\over
\gb{\ell|K|\ell}^{2+t}}{ \prod_{j=1}^{n+k} \vev{\ell|R_j
(K-s\eta)|\ell}\over \vev{\ell|\eta K|\ell}^n \prod_{p=1}^k
\vev{\ell| Q_p(K-s\eta)|\ell}}~.~~~\Label{Bnt}\eea
Bubble contributions in the cut integral (\ref{I-inte}) appear only if $n\geq 0$.
In the case that
$n=0$, it is easy to see that we can simply set $s=0$.  Then, $\vev{\ell|Q_p
K|\ell}=-\sqrt{1-u}\vev{\ell|q_p K|\ell}$, and $\vev{\ell|R_j
K|\ell}=-\sqrt{1-u}\vev{\ell|p_j K|\ell}$, so the $u$-dependent factor $\sqrt{1-u}$ cancels out of the numerator and denominator.

Our strategy is to decompose  $B_{n,t}(s)$ as a power series in $\sqrt{1-u}$.
 Then we will show that the
terms with odd powers of $\sqrt{1-u}$ correspond to spurious terms
discussed by OPP.  Therefore, they will vanish upon integration, and
we will be left with only even powers of  $\sqrt{1-u}$, i.e. a polynomial in $u$.

Note that  when we apply the OPP results, we are dealing only with the four-dimensional momentum $q$ (or $\W \ell$ in our notation), so we do not
need any parts of the OPP formulas involving the extra-dimensional variable $\W q^2$. Also, in our case we have
$p_0=0$, and the mass $m_i^2$ should be shifted to $m_i^2+\mu^2$.

We  emphasize one point which is crucial for our proof:
{\sl the one-to-one correspondence between the form (\ref{I-inte})
and the form (\ref{gen-n}) in $D=4$ dimensions}. That is,
every factor $-2\W\ell\cdot P_j$ in (\ref{I-inte})  corresponds to a factor $\vev{\ell|R_j^{(4D)} (K-s\eta)|\ell}$
in (\ref{gen-n}), and vice versa. It is very important that since now
we are in pure 4D, the $R_j^{(4D)}= -P_j$, i.e.,
$R_j^{(4D)}=R_j(u=0)$. Similarly for the factor $\vev{\ell|
Q_p(u=0)(K-s\eta)|\ell}$ in (\ref{gen-n}) and factor $(p-K_j)^2$ in
(\ref{I-inte}). In the following proof, we go back and
forth freely between these two forms.

%%%%%%%%%%%%%%%%%%%
\subsection{Reducing the number of propagators}
%%%%%%%%%%%%%%%%%%%%

The spurious terms of OPP have at most four propagator factors in the denominator.  In order to make use of their results, we must begin by reducing our (arbitrary) number of propagators to four or fewer.  In our formalism, the corresponding condition on (\ref{Bnt}) is that $k\leq 2$, because we have a unitarity cut.\\  We perform the reduction in (at most) two steps: first, from $k \geq 4 $ to $k\leq 3$, and then from $k=3$ to $k \leq 2$. \\

{\bf Reducing from $k \geq 4$ to $k\leq 3$:}\\

If $k\geq 4$, then there are at least 4 $Q_i$'s in the denominator and at least one $R$ in the numerator of (\ref{Bnt}).  Therefore, we can expand the vector $R$ in the basis of the $Q_i$, as follows.
\bea
R=\sum_i x_i Q_i ~~~\Label{opp-red43}
\eea
With this expansion, the original term can be expressed as a sum of four others, in each of which there is a cancellation between numerator and denominator, reducing $k$ to at most 3.  Of course, we must be sure that the coefficients $x_i$ are independent of $u$.  To see this, expand $Q_i$ and $R$ as in (\ref{Q-q-def}) and (\ref{R-p-def}) by writing $Q_i=-(\sqrt{1-u}) q_i+\a_i K$  and $R=-(\sqrt{1-u}) p+\b K$.  Then, we find that (\ref{opp-red43}) becomes
\bea  p=\sum_{i=1}^4 x_i q_i,~~~~\sum_i x_i \a_i=\b
\eea
Here it is clear that the solutions $x_i$ are independent of $u$.
Note that the equation $p=\sum_{i=1}^4 x_i q_i$ has only three
independent components, because the vectors $q_i$ span the
3-dimensional space orthogonal to $K$. Thus we have four equations
giving a unique solution of $x_i$.
\\

{\bf Reducing from $k=3$ to $k\leq 2$:}\\

Now we reduce further, from $k=3$ to $k\leq 2$. Since $k=3$ (and
we know $n \geq 0$, because we are discussing bubbles)
there is more than one $R$ in the numerator.
Taking any one of the $R$, we expand
\bean P=y_K K+ y_1 K_1+ y_2 K_2+y_3 K_3. \eean
Then,
\bean R &= & -\sqrt{1-u} \left( P - {P\cdot K\over K^2}K\right)+\b K
\\ & = & \sum_{i=1}^3 y_i \left\{-\sqrt{1-u} \left( K_i - {K_i\cdot K\over K^2}K\right)
+\a_i K\right\}+ (\b-\sum_{i=1}^3 y_i \a_i) K  \\
& = & \sum_{i=1}^3 y_i Q_i+ (\b-\sum_{i=1}^3 y_i \a_i) K \eean
Substituting this expansion into the numerator of (\ref{Bnt}), we obtain
\bean {\vev{\ell|R (K-s\eta)|\ell} \over \prod_{i=1}^3 \vev{\ell|Q_i
(K-s\eta)|\ell}}=\sum_{t=1}^3 { y_i\over \prod_{i=1,i\neq t}^3
\vev{\ell|Q_i (K-s\eta)|\ell}}+  (\b-\sum_{i=1}^3 y_i
\a_i){s\vev{\ell|\eta K|\ell} \over \prod_{i=1}^3 \vev{\ell|Q_i
(K-s\eta)|\ell}}\eean
The first three terms fall into the case $k=2$. The last term, with
the factor $s\vev{\ell|\eta K|\ell}$ in the numerator, still has
$k=3$, but we see from comparison with (\ref{Bnt}) that we have
effectively reduced $n$ by one. \footnote{There is another way to see
this point. The presence of a term $\vev{\ell|\eta K|\ell}$ implies that there is a
factor $(-2\W\ell\cdot K)$ in the form (\ref{I-inte}). By the
delta-function condition from the 4-dimensional unitarity cut,
this factor is
equivalent to $ K^2$, so we have reduced $n$ by one.} Repeating the
reduction on the last term, $n$ times, we arrive at a term with
$n=0$. As we discussed in the paragraph following (\ref{Bnt}), such
a term is independent of $u$.

Having accomplished the reduction of our proof to the case $k\leq 2$, we proceed to apply the results of OPP in a case-by-case analysis for $k=0,1,2$.

%%%%%%%%%%%%%%%%%%%
\subsection{Case-by-case analysis}
%%%%%%%%%%%%%%%%%%%

We now analyze each of the cases $k=0,1,2$ in turn, rearranging our integrand (\ref{Bnt}) so that the terms with odd powers of $\sqrt{1-u}$ take the form of the spurious terms of OPP \cite{Ossola:2006us}, which were proven there to vanish upon integration.\\

{\bf The case  $k=0$:}\\

We apply the OPP result directly and make use of their notation.
Recall that we always have $p_0=0$. Use (\ref{R-K-def}), i.e.
$R_j(u)=-\sqrt{1-u} \left( P_j - {P_j\cdot K\over K^2}K\right)+\b_j K$,
and expand $P_j$ as follows (see (2.23) of \cite{Ossola:2006us}):
\bea P_j= y_1 K+ y_n n+ y_7 \ell_7+y_8\ell_8.
\eea
 More concretely,
\bea K & = & \ell_5 +2 \ell_6 \\
 \ell_5 & = & K-{K^2\over 2K\cdot\eta}
\eta,~~~\ell_6={K^2\over 4K\cdot\eta}\eta \\
\ell_7& = & \la_{\ell_5}\W \la_{\ell_6},~~~~\ell_8= \la_{\ell_6}\W
\la_{\ell_5},~~~~n=\ell_5-2 \ell_6\eea
Here $\eta$ is the same null vector $\eta$, chosen arbitrarily, that we used inside $B_{n,t}(s)$. We see immediately that
\bea R_j(u)=
-(\b_j-y_n\b_n-y_{\ell_7}\b_{\ell_7}-y_{\ell_8}\b_{\ell_7}) R_K+ y_n
R_n+ y_7 R_{\ell_7}+y_8 R_{\ell_8}~~~\Label{opp-k0-exp} \eea
where $R_{\ell_7}=-\sqrt{1-u} \left( \ell_7 - {\ell_7\cdot K\over
K^2}K\right)-{\ell_7\cdot K\over K^2}K$, and the other three vectors are defined similarly from (\ref{R-p-def}).
After accounting for orthogonality properties,
 \bean n\cdot K=\ell_7\cdot K=\ell_8\cdot
K=0\eean
we can see specifically how the $u$-dependence enters:
\bea R_K=-K,~~~~R_{n}=-(\sqrt{1-u})n,~~~R_{\ell_7}=-(\sqrt{1-u})
\ell_7,~~~R_{\ell_8}=-(\sqrt{1-u}) \ell_8.  \eea

Now we use the one-to-one correspondence between form
(\ref{I-inte}) and form (\ref{Bnt}), that we emphasized at the beginning
of this section.  In the present analysis of a term of type (\ref{Bnt})
with $k=0$, and a numerator factor with $R_{\ell_7}$, the
corresponding term in (\ref{I-inte}) will have  the factor
 $-2\W\ell\cdot \ell_7$.
If we expand every $R_j$ according to (\ref{opp-k0-exp}),
then our general term is of the following form:
\bea (-2\W\ell\cdot K)^{s_K}(-2\W\ell\cdot \ell_7)^{s_7}
(-2\W\ell\cdot\ell_8)^{s_8}(-2\W\ell\cdot n)^{s_n}.~~~\Label{k=0-exp}
\eea
We have introduced integers $s_i$ to denote the powers.
The $u$-dependence of such a term is precisely the factor $\sqrt{1-u}^{s_7+s_8+s_n}$.

%Now we analyze each factor. First by our double
%cut we can see that $(-2\W\ell\cdot K)=K^2$, i.e., the role of this
%factor is to reduce the $n$ as we have discussed in previous
%subsection.
We need to show
that if $s_7+s_8+s_n$ is odd, then the term is spurious in the sense of OPP \cite{Ossola:2006us}.  First of all, the unitarity cut condition means we can replace $(-2\W\ell\cdot K) \to K^2$, so the value of $n$ is effectively reduced by one, and we can ignore that factor for the rest of the proof.
 If either $s_7=0$ or $s_8=0$, then we see immediately from
 \cite{Ossola:2006us}-(2.29) that the term is spurious. When both $s_7, s_8$
are nonzero, we use the expression \cite{Ossola:2006us}-(2.33) to reduce to the
case of \cite{Ossola:2006us}-(2.29). If $s_7\neq s_8$, the conclusion is obvious. But
if $s_7=s_8$, then $s_n$ is odd, and so, after applying \cite{Ossola:2006us}-(2.33), $2i+s_n$ is still an odd power, and
we can again conclude with \cite{Ossola:2006us}-(2.29) that the term is spurious.  Finally, we must account for the first two terms in \cite{Ossola:2006us}-(2.33).
The first is $\sum_{i=0}^1 {\cal O}(D_i)$, which is zero  by the unitarity cut
condition. The second term is ${\cal O}(\W q^2)$, which is zero since our present analysis is purely four-dimensional, as we remarked at the beginning of this section.
\\

{\bf The case $k=1$:}\\

We continue using the notation of \cite{Ossola:2006us}, and also its discussion of
3-point like spurious terms.
Now we use the following expansion involving the single vector $K_i$:
\bea P_j= y_K K+ y_i  K_i + y_3 \ell_3+y_4\ell_4.
\eea
The vectors $\ell_3,\ell_4$ are defined as $\ell_3=\la_{\ell_1}\W \la_{\ell_2}$, $\ell_4=\la_{\ell_2}\W
\la_{\ell_1}$, where $\ell_1, \ell_2$ are constructed from $K,K_i$.
Then,
\bean R_j(u) & = &  -\sqrt{1-u} \left( P_j - {P_j\cdot K\over
K^2}K\right) +\b_j  K \\ & = & -(\b_j-y_i\a_i-y_3
\b_{\ell_3}-y_4\b_{\ell_4}) R_K+ y_i Q_i+ y_3 R_{\ell_3}+y_4
R_{\ell_4}\eean

Now we substitute this expansion into  $\vev{\ell|R_j(u)
(K-s\eta)|\ell}$. The term with $Q_i$ cancels the a factor in the
denominator, returning us to the case of $k=0$, which we have
already addressed. The term with $R_K$ reduces $n$ by one, as
discussed above. For the remaining two terms,  we use $\ell_3\cdot
K=\ell_4\cdot K=0$ to write $R_{\ell_3}=-(\sqrt{1-u})\ell_3$ and
$R_{\ell_4}=-(\sqrt{1-u}) \ell_4$, just as in the case of $k=0$.

However, unlike the $k=0$ case, when we put $R_{\ell_3}$ and
$R_{\ell_4}$ back in (\ref{Bnt}),  the power of
$\sqrt{1-u}$ is not always given by the power of $R_{\ell_3}$ and
$R_{\ell_4}$, so we must be careful. Let us
consider the separate cases for each term in the expansion.
\begin{itemize}

\item (a) If the term contains neither $R_{\ell_3}$ nor
$R_{\ell_4}$, then either $Q_i$ effectively reduces $k=1$ to $k=0$, or $R_K$
reduces $n$ to $n=0$ in $n$ steps. Either way, we know from previous analysis that the odd powers of $\sqrt{1-u}$ drop out.

\item (b) If the term contains $R_{\ell_3}$ or $R_{\ell_4}$, but not both, i.e.,
\bean {(-2\W\ell\cdot \ell_3)^a\over (\W\ell-K_i)^2-\mu^2}~~~or~~{
(-2\W\ell\cdot \ell_4)^b\over (\W\ell-K_i)^2-\mu^2},~~~a, b\neq
0\eean
by \cite{Ossola:2006us}-(2.20), the contribution is zero.

\item (c) If the term contains both  $R_{\ell_3}$ and
$R_{\ell_4}$,i.e.,
\bean {(-2\W\ell\cdot \ell_3)^a (-2\W\ell\cdot \ell_4)^b\over
(\W\ell-K_i)^2-\mu^2},~~~a,b\neq 0\eean
then we need to use the first equation of \cite{Ossola:2006us}-(2.15) to reduce
the pair. There are three terms on the right hand side of the first
equation (remembering that the ${\cal O}(\W q^2)$ does not exist in
our case). The first two terms reduce $n$ by two (notice that $F$ depends on $\mu^2$ through the mass), and the third
term reduces $k=1$ to $k=0$. By this manipulation, we reduce case (c)
to either case (a) or case (b).

\end{itemize}

\vskip0.1in

{\bf The case $k=2$:}\\

We use the same expansion of $R_j$ as in the $k=1$ case, and perform a similar analysis. Factors with $R_K$
factor effectively reduce $n$ by one.
Factors with $Q_1$ reduce $k=2$ to $k=1$. For $R_{\ell_3}$ and $R_{\ell_4}$, we need
to use equation \cite{Ossola:2006us}-(2.15) to simplify further. Similar to the case
$k=1$, we have following three cases:

\begin{itemize}

\item (a)  If the term contains neither $R_{\ell_3}$ nor
$R_{\ell_4}$, then either $Q_i$ reduces $k=2$ to $k=1$, or $R_K$
reduces $n$ to $n=0$ in $n$ steps.  Either way, we know from previous analysis that the odd powers of $\sqrt{1-u}$ drop out.

\item (b)  If the term contains $R_{\ell_3}$ or $R_{\ell_4}$, but not both,
i.e.,
\bean {(-2\W\ell\cdot \ell_3)^a\over
((\W\ell-K_i)^2-\mu^2)((\W\ell-K_j)^2-\mu^2)}~~~or~~{ (-2\W\ell\cdot
\ell_4)^b\over ((\W\ell-K_i)^2-\mu^2)((\W\ell-K_j)^2-\mu^2)},~~~a,
b\neq 0\eean
then we apply the second equation of \cite{Ossola:2006us}-(2.15)
repeatedly until we reach the form \cite{Ossola:2006us}-(2.18). There are three terms on the
right-hand side of \cite{Ossola:2006us}-(2.18). The first term will depend on $u$
polynomially through the mass, while the third term will reduce
$k=2$ to $k=1$. The second term
is the spurious term, which gives zero contribution.

\item (c)  If the term contains both  $R_{\ell_3}$ and
$R_{\ell_4}$,i.e.,
\bean {(-2\W\ell\cdot \ell_3)^a (-2\W\ell\cdot \ell_4)^b\over
((\W\ell-K_i)^2-\mu^2)((\W\ell-K_j)^2-\mu^2)},~~~a,b\neq 0\eean
then we apply the first equation of \cite{Ossola:2006us}-(2.15) to reduce
the pair. Then we use the second equation of \cite{Ossola:2006us}-(2.15) and finally
reach the form of (2.18). The discussion of this case is parallel to
case (b).

\end{itemize}

We conclude that the bubble coefficients, as given in
(\ref{bub-exp}), are indeed polynomials in $u$.  Knowing this fact,
the degree of the polynomial can then be read off from the formulas of
Section 4; it is seen to be $[n/2]$.

%

%%%%%%%%%%%%%%%%%%%%%%%%%%%%%
\section{\label{ap:pentagon}Pentagon double cut}
%%%%%%%%%%%%%%%%%%%%%%%%%%%%%

The double cut of a pentagon integral, defined according to (\ref{I-inte}) as
\bea  C[I_5(K;K_1,K_2,K_3)] &=&  \int d^{4-2\eps} p~ (\mu^2)\frac{ 1 }{\prod_{j=1}^3 (p-K_j)^2} \delta(p^2)
\delta((p-K)^2)
\eea
 is given by the following expression \cite{Anastasiou:2006gt}:
\bea C[I_5(K;K_1,K_2,K_3)] &=& -\int_0^1 du~ u^{-1-\eps}
 {\sqrt{1-u} \over (K^2)^2}~~~\label{Pentagon-gen} \\ &  &
\left({ S[Q_3, Q_2, Q_1,K]\over
4\sqrt{(Q_3\cdot Q_2)^2 -Q_3^2 Q_2^2}} \ln{  Q_3 \cdot Q_2 -
\sqrt{  (Q_3\cdot Q_2)^2 -Q_3^2 Q_2^2}\over  Q_3 \cdot Q_2 +
\sqrt{  (Q_3\cdot Q_2)^2 -Q_3^2 Q_2^2}}\right. \\ & & + { S[Q_3,
Q_1, Q_2,K]\over 4\sqrt{(Q_3\cdot Q_1)^2 -Q_3^2 Q_1^2}} \ln{
Q_3 \cdot Q_1 - \sqrt{  (Q_3\cdot Q_1)^2 -Q_3^2 Q_1^2}\over  Q_3
\cdot Q_1 + \sqrt{  (Q_3\cdot Q_1)^2 -Q_3^2 Q_1^2}}  \\ & &
\left.+ { S[Q_2, Q_1, Q_3,K]\over 4\sqrt{(Q_2\cdot Q_1)^2
-Q_2^2 Q_1^2}} \ln{  Q_2 \cdot Q_1 - \sqrt{  (Q_2\cdot Q_1)^2
-Q_2^2 Q_1^2}\over  Q_2 \cdot Q_1 + \sqrt{ (Q_2\cdot Q_1)^2
-Q_2^2 Q_1^2}}\right),
\nonumber \eea
where $S[Q_i, Q_j, Q_k,K]$ was defined to be
\bea
& & S[Q_i, Q_j, Q_k,K]  =  {T_1\over T_2},~~~\label{Func-S-sec}
\eea
with
\bea
T_1 = -8 \det \left( \begin{array}{lcr} K \cdot Q_k & Q_i \cdot K & Q_j \cdot K\\
Q_i \cdot Q_k & Q_i^2 & Q_i \cdot Q_j \\ Q_j \cdot Q_k & Q_i \cdot Q_j &
Q_j^2
\end{array} \right); ~~~~
T_2 = -4 \det \left( \begin{array}{lcr} Q_k^2 & Q_i \cdot Q_k & Q_j \cdot Q_k\\
Q_i \cdot Q_k & Q_i^2 & Q_i \cdot Q_j \\ Q_j \cdot Q_k & Q_i \cdot Q_j &
Q_j^2
\end{array} \right).~~~~\label{T1-T2-sec}
\eea
Here we rewrite (\ref{Func-S-sec}) so that the $u$-dependence becomes transparent.
We need to define a few auxiliary quantities. In terms of a particular
matrix
denoted by $S$,
\bea S \equiv \left( \begin{array}{ccccc} 0 & K^2 & K_1^2 & K_2^2 & K_3^2
\\ K^2 & 0 & (K_1-K)^2 & (K_2-K)^2 & (K_3-K)^2 \\ K_1^2 & (K_1-K)^2 &
0 & ( K_2-K_1)^2 & (K_3-K_1)^2 \\ K_2^2 & (K_2-K)^2
 & (K_2-K_1)^2 & 0 & (K_3-K_2)^2 \\ K_3^2 & (K_3-K)^2 & (K_3-K_1)^2 &
 (K_3-K_2)^2 & 0 \end{array}\right),\eea
we define
\bea A[K_1; K_2,K_3,K] & = &
 -{\rm det}\left( \begin{array}{ccccc}
  ~~0~~ & ~~K_2^2~~ & ~~K_3^2~~ & ~~K^2~~ & ~~ K_1^2 ~~ \\  K_2^2 & 0 &
  (K_2-K_3)^2 & (K_2-K)^2  & (K_2-K_1)^2 \\ K_3^2 &
  (K_3-K_2)^2 & 0 & (K_3-K)^2  & (K_3-K_1)^2  \\ K^2 &
  (K-K_2)^2 & (K-K_3)^2  & 0 & (K-K_1)^2 \\ 1 & 1 & 1 & 1 & 1\end{array}\right), \\
B[K_1, K_2, K_3, K] &=&
{\rm det}(S)\sum_{i,j=1}^5 (S^{-1})_{ij}, \\
C[K_1, K_2, K_3, K] & = & 2 {\rm det}(S).
\eea
Then,
\bea  S[Q_2, Q_3, Q_1, K] & = & { 4 K^2 A[K_1; K_2, K_3, K]\over u
K^2 B[K_1,K_2, K_3, K] - C[K_1, K_2, K_3, K]}.~~~\Label{S-ABC}
\eea
Now it is evident that the numerator of (\ref{S-ABC}) is independent of $u$, and the denominator is linear in $u$.  Furthermore, $B[K_1,K_2, K_3, K]$ and $C[K_1, K_2, K_3, K]$ are totally symmetric in their arguments, indicating fundamental pentagon nature. The quantity $A[K_1; K_2, K_3, K]$ breaks this symmetry for the first argument, $K_1$, indicating that the corresponding propagator is the one that is eliminated in order to show up as part of a {\em box} coefficient.

%%%%%%%%%%%%%%%%%%%%%%%%%%%%%%%%%%%%%%%%%%%%%%%%%%%%

\end{document}